\documentclass[preprint,eqsecnum,aps,nofootinbib]{revtex4}
\usepackage{amsfonts,amsmath,amssymb,amsthm}
\usepackage{latexsym}
\usepackage{bbm,bm}
\usepackage{graphicx}

\newcommand{\ra}{\rangle}
\newcommand{\la}{\langle}

\renewcommand{\o}{\otimes}

\begin{document}

\title{Effect of Phase Factor in the Geometric Entanglement Measure of Three-Qubit States}

\author{Sayatnova Tamaryan}

\affiliation{Theory Department, Yerevan Physics Institute, Yerevan-36, 375036, Armenia}

\author{Hungsoo Kim}

\affiliation{The Institute of Basic Science, Kyungnam University, Masan, 631-701, Korea}

\author{Mu-Seong Kim, Kap Soo Jang, DaeKil Park}

\affiliation{Department of Physics, Kyungnam University, Masan, 631-701, Korea}

\begin{abstract}
Any pure three-qubit state is uniquely characterized by one phase and four positive 
parameters. The geometric measure of entanglement as a function of state parameters 
can have different expressions. Each of expressions has its own applicable domain 
and thus the whole state parameter space is divided into subspaces that are ranges 
of definition for corresponding expressions. The purpose of this paper is to examine 
the applicable domains for the most general qubit-interchange symmetric three-qubit
states. First, we compute the eigenvalues of the non-linear eigenvalue equations
and the nearest separable states for the permutation invariant 
three-qubit states with a fixed phase. Next, we compute the geometric entanglement 
measure, deduce the boundaries of all subspaces, 
and find allocations of highly and slightly entangled states. It is shown that 
there are three applicable domains when the phase factor is $\pi/2$ while other
cases have only two domains. The emergence of the three domains is due to the 
appearance of the additional W-state. We show that most
of highly entangled states reside near the boundaries of the domains and states 
located far from the boundaries become less-entangled and eventually go to the 
product states. The neighbors of W-state are generally more entangled than the 
neighbors of Greenberger-Horne-Zeilinger(GHZ) state from the aspect of the geometric 
measure. However, the range of the GHZ-neighbors is much more wider than the range 
of the W-neighbors.
\end{abstract}


\maketitle

\section{Introduction}

Entanglement is a property of quantum states that does not exist classically.
Two or more subsystems of a quantum system are said to be entangled if the state of the
entire system cannot be described in terms of a state for each of the
subsystems~\cite{wern-89}. This property of composite quantum systems, which exhibits
quantum correlations between subsystems, is a resource for many processes in quantum
information theory~\cite{ek-91,ben-wies,ben-per,ital-09}.
Since the profound measures of entanglement, i.e. the entanglement of formation and
distillation~\cite{ben-schum,ben-vinc,woot-98,vid-cir}, have not been properly
generalized to multiparticle systems, the study of quantifying multipartite entanglement
via other measures~\cite{vedr-plen,plen-vedr,mey-wall,grov-02,shap-bih} is a necessity.

The entanglement of a given pure state can be characterized by a
distance to the nearest unentangled state~\cite{shim-95}. A whole class of such entanglement
monotones, based on the Euclidean distance of a given multipartite state to the
nearest fully separable state, was constructed in Ref.\cite{barn-01}.
Subsequently, a geometrically motivated measure of entanglement, known as geometric measure,
was introduced by Wei and Goldbart~\cite{wei-03}. It is a decreasing function of the
maximal overlap $P_{max}$ and is suitable for any partite system regardless of its
dimensions. The maximal overlap has several different names and we 
list all of them for the completeness:
maximal probability of success~\cite{grov-02}, entanglement eigenvalue~\cite{wei-03},
injective tensor norm~\cite{wern-02}, the largest Schmidt coefficient~\cite{gsd-08} and
maximum singular value~\cite{sud-geom}.

The geometric measure has an advantage that it can be computed analytically for 
multi-parameter states. Recently, explicit expressions for the maximal overlap have 
been derived for three-\cite{wei-03,sud-geom,analytic,shared,3q} as well as for 
multi-qubit states~\cite{bih-04,local,guh-mix,toward}. It turned out that the maximal 
overlap, depending on coefficients of a quantum state in a computational basis, can take 
two different values. It is equal to either the square of the largest coefficient or 
the square of the circumradius of a cyclic polygon constructed by the coefficients of 
the quantum state. This means that the whole
parameter space is divided into two subspaces each of which has its own expression for the
geometric measure.

In spite of these achievements, still we lack sufficient knowledge to classify generic 
three-qubit pure states by the geometric measure. They have five local unitary(LU) 
invariants including four positive parameters and a gauge phase 
$\gamma$~\cite{gsd-08,acin,hig}. The maximal overlap of these states is not known yet. 
Only three-qubit states which are expressed as linear combinations of four(or less) 
orthogonal product states have been considered so far~\cite{shared}. In fact, all of 
these states have real coefficients because the phases of their coefficients can be 
eliminated by LU-transformations. Thus, the contribution of the gauge phase to the 
maximal overlap has remained a mystery. On the other hand, the most recent 
results~\cite{maximal} have shown that the gauge phase plays an important role. 
It parameterizes the family of maximally entangled states and identifies W-class pure 
states with the boundary of pure states.

In this paper we would like to take into complete account the effect of the gauge phase 
in the geometric measure of entanglement. We compute the maximal overlap as well as 
the nearest product states for a given value of the gauge phase. We will show in the 
following that depending on the phase factor $\gamma$ the whole parameter space is 
divided into the two or three domains, each of which has a particular expression for 
the geometric measure. In addition, we will show that most of highly entangled states 
reside near the boundaries of the domains. We will call these highly entangled states 
as GHZ-neighbors. The states located far from the boundaries become less-entangled and 
eventually go to the product states. But there is different 
kind of the highly entangled states. 
These states reside around W-states. We will call these highly entangled states as W-neighbors.
The W-neighbors are generally more entangled than the GHZ-neighbors from the aspect of 
the geometric measure. However, the range of the GHZ neighbors is much more wider than 
the range of the W-neighbors.

The paper is organized as follows. In section II following Ref.\cite{analytic} we transform
the nonlinear eigenvalue equations into the Lagrange multiplier equations. In section III
we solve the Lagrange multiplier equations analytically for $\gamma = 0$ and $\gamma = \pi/2$.
It turns out that both cases give five different eigenvalues. Also every eigenvalue has its
own available region in the parameter space. In section IV we compute the geometric measure
for $\gamma=0$ case. It turns out that two of the five eigenvalues contribute to the geometric
measure. This means that the whole parameter space is divided into two applicable domains.
In section V we compute the geometric measure for $\gamma = \pi/2$ case. It is shown that the
whole parameter space is divided into the three applicable domains. In section VI we compute
the eigenvalues and the geometric measure for $\gamma = \pi/4$ numerically. It is shown that
when $\gamma=\pi/4$, there are six different eigenvalues. However, only two eigenvalues 
contribute to the geometric measure. In section VI a brief conclusion is given. 
In appendix we have shown that Lagrange multiplier equations for arbitrary $\gamma$ provides 
a solution whose multiplier constant is zero.

\section{General Formalism}

In this section we clarify our notations, give necessary definitions, define three-qubit
symmetric states and transform nonlinear stationarity equations to a system of linear equations.

\subsection{Preliminaries}

The maximal overlap of $n$-qubit pure states is given by
\begin{equation}\label{injective}
P_{max}=\max_{q_1,q_2,...,q_n}|(\la q_1|\o\la q_2|\o\cdots\o\la q_n|)|\psi\ra|^2,
\end{equation}
where the maximization is performed over single qubit pure states. Constituents
$|q_1\ra$, $|q_2\ra$, ..., $|q_n\rangle$, the nearest product state from $|\psi\rangle$,
can be computed via the non-linear eigenvalue equations
\begin{equation}
\label{nonlinear-ev}
\langle q_1|\cdots\langle q_{n-1}|\psi\rangle=\mu_i |q_n\rangle, \;
\langle q_1|\cdots \langle q_{n-2}|\langle q_n|\psi\rangle = \mu_i |q_{n-1}\rangle,\;
\cdots,\;
\langle q_2| \cdots \langle q_n|\psi\rangle = \mu_i |q_1\rangle,
\end{equation}
where $\mu_i$'s are the eigenvalues of Eq.(\ref{nonlinear-ev}). Then
the geometric measure $G$ of the quantum state $|\psi\rangle$ is defined as
$G(\psi) = 1 - P_{max}$, where $P_{max} = \max (\mu_i^2)$.

For simplicity, we take a quantum states which
possess a permutational symmetry~\cite{enk-09,guhn-gez,wei-sev}. These states have three
independent  parameters and, through an appropriate LU transformations, can be 
brought into
the symmetric form~\cite{gsd-08}
\begin{equation}
\label{symmetric}
|\psi\ra=g|000\ra+t|011\ra+t|101\ra+t|110\ra+ e^{i\gamma}h|111\ra,
\end{equation}
where we follow the notation of Ref.\cite{maximal}. In above equation all coefficients
$g$, $h$ and $t$ are positive and satisfy the normalization condition
$g^2 + 3 t^2 + h^2 = 1$. The phase $\gamma$ has the period $\pi$  and ranges within
the interval $-\pi/2 \leq \gamma \leq \pi/2$. Note that Eq.(\ref{symmetric}) is not a
Schmidt decomposition for $|\psi\ra$ since the Schmidt normal form imposes additional
conditions(namely, a lower bound on $g$) on state parameters. We would like to abandon these additional constraints
and apply the general method proposed in Ref.\cite{analytic} to symmetric
states Eq.(\ref{symmetric}).

\subsection{Modified stationarity equations}

In this subsection we would like to present the method for solving stationarity equations
for the quantum state given in Eq.(\ref{symmetric}). In the case of three-qubit pure states 
the method developed in Ref.\cite{analytic} transforms the system of nonlinear
equations to a system of linear equations. In spite of this essential simplification,
it is impossible to get analytic expressions for generic three-qubit states since the
solution of the linear eigenvalue equations reduces to the root finding for a couple of
algebraic equations of degree six~\cite{shared}. However, the permutation symmetry of 
$|\psi\rangle$ reduces this pair of algebraic equations to a single algebraic equation 
of degree six. Furthermore, there is a solution which holds for all values of state
parameters~\cite{maximal}. The separation of this global solution allows us to
solve explicitly the eigenvalue equations for $\gamma=0$ and $\gamma=\pi/2$ and leads us to
a quartic equation for remaining cases. The quartic is the highest order polynomial equation 
that can be solved by radicals in the general case. But expressions for roots are impractical 
and we will carry out numerical analysis instead.

The method enables us to express eigenvalues $\mu^2$  via the reduced densities
$\rho^{A},\,\rho^{B}$ and $\rho^{AB}$ of qubits A and B in a form:
\begin{equation}
\label{eigenvalue-1}
\mu^2 = \frac{1}{4} \max_{|{\bm s}_1| = |{\bm s}_2| = 1}
\left( 1 + {\bm r}_1 \cdot {\bm s}_1 +  {\bm r}_2 \cdot {\bm s}_2
+ G_{ij} s_{1i} s_{2 j} \right),
\end{equation}
where
\begin{equation}
\label{def-2-1}
{\bm r}_1 = \mbox{Tr} \left(\rho^A {\bm \sigma} \right),  \hspace{1.0cm}
{\bm r}_2 = \mbox{Tr} \left(\rho^B {\bm \sigma} \right),  \hspace{1.0cm}
G_{ij} = \mbox{Tr} \left( \rho^{AB} \sigma_i \otimes \sigma_j \right)
\end{equation}
and $\sigma_i$'s are Pauli matrices. Explicit calculation shows
\begin{eqnarray}
\label{quant-2-1}
& & {\bm r} \equiv {\bm r}_1 = {\bm r}_2 = (2ht\cos\gamma, 2ht\sin\gamma, g^2-h^2-t^2)
                                                                          \\  \nonumber
& & G_{ij} = \left(                \begin{array}{ccc}
                      2 t (g + t)   &    0    & -2ht\cos\gamma    \\
                      0      &     -2 t (g - t)   & -2ht\sin\gamma   \\
                      -2ht\cos\gamma & -2ht\sin\gamma & g^2+h^2-t^2
                                   \end{array}                    \right).
\end{eqnarray}
It is worthwhile noting that ${\bm r}_1$ is identical with ${\bm r}_2$ and $G_{ij}$ is a
symmetric matrix. These properties arise due to the fact that we have chosen the
symmetric state in Eq.(\ref{symmetric}) under the qubit-exchange.
As will be shown in the following these properties drastically simplify the calculation procedure.
Since ${\bm r}_1$, ${\bm r}_2$ and $G_{ij}$ are explicitly derived, the eigenvalues $\mu^2$
can be computed if ${\bm s}_1$ and ${\bm s}_2$ are known. Due to the maximization in
Eq.(\ref{eigenvalue-1}) these vectors can be computed by solving the Lagrange multiplier
equations:
\begin{equation}
\label{lagrange-2-1}
{\bm r}_1 + G {\bm s}_2 = \lambda_1 {\bm s}_1       \hspace{1.0cm}
{\bm r}_2 + G^T {\bm s}_1 = \lambda_2 {\bm s}_2
\end{equation}
where the superscript $T$ stands for transpose and $\lambda_i$'s are the Lagrange multiplier
constants. From the properties ${\bm r}_1 = {\bm r}_2$ and $G_{ij} = G_{ji}$
Eq.(\ref{lagrange-2-1}) can be reduced to a single equation
\begin{equation}
\label{lagrange-2-2}
{\bm r} + G {\bm s} = \lambda {\bm s}
\end{equation}
where $\lambda \equiv \lambda_1 = \lambda_2$ and ${\bm s} \equiv {\bm s}_1 = {\bm s}_2$.
Letting
\begin{equation}
\label{para-2-1}
{\bm s} = (\sin \theta \cos \varphi, \sin \theta \sin \varphi, \cos \theta),
\end{equation}
Eq.(\ref{lagrange-2-2}) reduces to
\begin{subequations}
\label{main-2}
\begin{equation}
\label{main-2x}
2ht \cos \gamma + 2 t (g + t) \sin \theta \cos \varphi - 2 h t \cos \gamma \cos \theta
       = \lambda \sin \theta \cos \varphi
\end{equation}
\begin{equation}
\label{main-2y}
2 h t \sin \gamma - 2 t (g-t) \sin\theta \sin \varphi - 2 h t \sin\gamma \cos \theta
       = \lambda \sin \theta \sin \varphi
\end{equation}
\begin{equation}
\label{main-2z}
(g^2 - t^2) (1 + \cos \theta) - h^2 (1 - \cos \theta) - 2ht \cos\gamma \sin\theta\cos \varphi
- 2ht \sin\gamma \sin\theta \sin\varphi = \lambda \cos \theta.
\end{equation}
\end{subequations}
Solving $\theta$, $\varphi$ and $\lambda$ from Eq.(\ref{main-2}), one can compute the
eigenvalues for the symmetric canonical state (\ref{symmetric}) by inserting
the solutions into Eq.(\ref{eigenvalue-1}). In the next section we will solve analytically 
Eq.(\ref{main-2}) at the particular phases $\gamma = 0$ and $\gamma = \pi/2$. By making
use of the solutions we will compute $\mu_i$ and $P_{max} = \max (\mu_i^2)$ for the 
corresponding quantum states. 

\section{Eigenvalues}

In this section Eq.(\ref{main-2}) will be solved at $\gamma = 0$ and $\pi/2$
separately. Since numerical calculation is needed to analyze the $\gamma=\pi/4$ case,
we deal with this case in different section (see section VI).

\subsection{$\gamma = 0$ case}
For this case Eq.(\ref{main-2}) reduces to
\begin{subequations}
\label{ga-0}
\begin{equation}
\label{ga-0x}
2 t(g+t)\sin\theta\cos\varphi +2ht(1-\cos\theta)=\lambda\sin\theta\cos\varphi
\end{equation}
\begin{equation}
\label{ga-0y}
-2 t(g-t)\sin\theta\sin\varphi = \lambda\sin\theta\sin\varphi
\end{equation}
\begin{equation}
\label{ga-0z}
(g^2-t^2)(1+\cos\theta)-h^2(1-\cos\theta) -2ht\sin\theta\cos\varphi = \lambda \cos\theta.
\end{equation}
\end{subequations}
Eq.(\ref{ga-0y}) implies that the solutions for the $\gamma = 0$ case are categorized by
$\theta=0$, $\varphi=0$, $\varphi=\pi$ and $\lambda = -2t(g-t)$\footnote{The case
$\theta = \pi$ can be excluded by Eq.(\ref{ga-0x}).}.

\subsubsection{$\theta=0$ case}
When $\theta=0$, Eq.(\ref{ga-0x}) and Eq.(\ref{ga-0y}) are automatically solved, and
Eq.(\ref{ga-0z}) gives
\begin{equation}
\label{lam-1}
\lambda = 2 (g^2 - t^2).
\end{equation}
Now ${\bm s} = (0, 0, 1)$ and Eq.(\ref{eigenvalue-1}) together with the normalization
condition $g^2 + 3 t^2 + h^2 = 1$ gives the eigenvalue
\begin{equation}
\label{ev-0-1}
\mu_P^2 = g^2.
\end{equation}

\subsubsection{$\varphi = 0$ case}
For this case Eq.(\ref{ga-0y}) is automatically solved and the remaining equations
are
\begin{subequations}
\label{couple-1}
\begin{equation}
\label{couple-1x}
2 t(g+t)\sin\theta + 2ht(1-\cos\theta)=\lambda\sin\theta
\end{equation}
\begin{equation}
\label{couple-1z}
(g^2-t^2)(1+\cos\theta)-h^2(1-\cos\theta) -2ht\sin\theta = \lambda \cos \theta.
\end{equation}
\end{subequations}
Since $\sin (\theta / 2) \neq 0$, Eq.(\ref{couple-1x}) reduces to
\begin{equation}
\label{lam-2}
\lambda = 2 h t z + 2 t^2 + 2 t g
\end{equation}
where $z = \tan (\theta/2)$. Inserting Eq.(\ref{lam-2}) into Eq.(\ref{couple-1z}), one can
derive an equation
\begin{equation}
\label{factor-1}
(h z + g + t)(t z^2 -h z + g- 2 t)=0.
\end{equation}
Eq.(\ref{factor-1}) implies that the $\varphi=0$ case is also categorized again by
following three cases:
\begin{equation}
\label{cate-1}
z = -\frac{g + t}{h}, \hspace{.5cm} \frac{r_+}{2 t}, \hspace{.5cm} \frac{r_-}{2 t}
\end{equation}
where
\begin{equation}
\label{rpm}
r_{\pm} = h \pm \sqrt{h^2 + 4 t (2 t - g)}.
\end{equation}

First, let us consider the case of $z = -(g+t)/h$. In this case Eq.(\ref{lam-2}) gives
\begin{equation}
\label{lam-3}
\lambda = 0.
\end{equation}
Since, in this case,
\begin{equation}
\label{s-0-1}
s_x = \sin \theta = -\frac{2 h (g+t)}{h^2 + (g+t)^2}, \hspace{1.0cm} s_y = 0, \hspace{1.0cm}
s_z = \frac{h^2 - (g+t)^2}{h^2 + (g+t)^2},
\end{equation}
it is straightforward to compute the eigenvalues for this case, which is
\begin{equation}
\label{ev-0-2}
\mu_1^2 = \frac{g^2 h^2 + t^2 (g+t)^2}{h^2 + (g+t)^2}.
\end{equation}

Next, let us consider the case of $z = r_{\pm}/2t$ simultaneously. In these cases
Eq.(\ref{lam-2}) gives
\begin{equation}
\label{lam-4}
\lambda = h r_{\pm} + 2 t (g+t).
\end{equation}
Since, in these cases,
\begin{equation}
\label{s-0-2}
s_x = \frac{4 t r_{\pm}}{r_{\pm}^2 + 4 t^2}, \hspace{1.0cm} s_y=0, \hspace{1.0cm}
s_z = - \frac{r_{\pm}^2 - 4 t^2}{r_{\pm}^2 + 4 t^2},
\end{equation}
one can show directly that the eigenvalues are
\begin{equation}
\label{ev-0-3}
\mu_{\pm}^2 = \frac{(h r_{\pm} + 4 t^2)^2}{r_{\pm}^2 + 4 t^2}.
\end{equation}
Since $z=\tan (\theta/2)$ should be real, the eigenvalues $\mu_{\pm}^2$ are available only
when
\begin{equation}
\label{avail-0-1}
g \leq 2 t + \frac{h^2}{4 t}.
\end{equation}

\subsubsection{$\varphi = \pi$ case}
For this case Eq.(\ref{ga-0y}) is automatically solved and the remaining equations are
\begin{subequations}
\label{couple-2}
\begin{equation}
\label{couple-2x}
-2 t(g+t)\sin\theta + 2ht(1-\cos\theta)= -\lambda\sin\theta
\end{equation}
\begin{equation}
\label{couple-2z}
(g^2-t^2)(1+\cos\theta)-h^2(1-\cos\theta) + 2ht\sin\theta = \lambda \cos \theta.
\end{equation}
\end{subequations}
Since Eq.(\ref{couple-2}) can be derived from Eq.(\ref{couple-1}) by changing
$\theta \rightarrow -\theta$, the solutions for this case are also categorized by
\begin{equation}
\label{cate-2}
z = \frac{g + t}{h}, \hspace{.5cm} -\frac{r_+}{2 t}, \hspace{.5cm} -\frac{r_-}{2 t}.
\end{equation}
Since Eq.(\ref{couple-2x}) reduces to
\begin{equation}
\label{lam-5}
\lambda = -2 h t z + 2 t^2 + 2 t g,
\end{equation}
comparison of Eq.(\ref{lam-5}) with Eq.(\ref{lam-2}) shows that the Lagrange multiplier
constant $\lambda$ is same with the case of $\varphi = 0$. Since, furthermore,
$s_x = \sin \theta \cos \varphi$ and $s_z = \cos \theta$ are invariant under
$\theta \rightarrow -\theta$ and $\varphi=0 \rightarrow \varphi = \pi$, this fact implies
that the eigenvalues for this case are exactly same with those for $\varphi=0$ case.

\subsubsection{$\lambda=2t^2-2gt$ case}
For this case Eq.(\ref{ga-0y}) is automatically solved and the remaining equations are
\begin{subequations}
\label{couple-3}
\begin{equation}
\label{couple-3x}
2 t(g + t)\sin\theta\cos\varphi + 2 h t (1 - \cos \theta) = -2t(g - t)\sin\theta\cos\varphi
\end{equation}
\begin{equation}
\label{couple-3z}
(g^2-t^2)(1+\cos\theta)-h^2 (1-\cos\theta)-2ht \sin\theta\cos \varphi = -2t(g-t)\cos\theta.
\end{equation}
\end{subequations}
Since Eq.(\ref{couple-3x}) gives a relation
\begin{equation}
\label{phi-t-1}
\cos\varphi = -\frac{h}{2 g} \frac{1 - \cos\theta}{\sin\theta},
\end{equation}
combining Eq.(\ref{couple-3z}) and Eq.(\ref{phi-t-1}) enables us to express $\cos \theta$ and
$\sin \theta$ as
\begin{equation}
\label{t-t-1}
\cos\theta = -\frac{g^2-h^2+gt}{g^2+h^2+3gt} \hspace{1.0cm}
\sin\theta=\pm \frac{\sqrt{4g(g+2t)(h^2 + gt)}}{g^2+h^2+3gt}.
\end{equation}
For a time being we choose the upper sign in $\sin\theta$. Then, Eq.(\ref{phi-t-1}) reduces to
\begin{equation}
\label{phi-2}
\cos\varphi = \frac{h}{2} \sqrt{\frac{g+2t}{g(h^2+gt)}}.
\end{equation}
At this stage it is worthwhile noting that the eigenvalue in this case is available when
\begin{equation}
\label{avail-0-2}
(3 g - 2 t) h^2 + 4 g^2 t \geq 0
\end{equation}
because of $-1 \leq \cos \varphi \leq 1$. Of course, the corresponding $\sin\varphi$ is
\begin{equation}
\label{sin-psi-1}
\sin \varphi = \pm \sqrt{\frac{3 g h^2+4g^2t-2 h^2t}{4g(h^2+gt)}}.
\end{equation}
Again we choose the upper sign in $\sin\varphi$. Then, it is straightforward to compute
${\bm s}$, whose components are
\begin{equation}
\label{s-0-3}
s_x = -\frac{h(g+2t)}{g^2+h^2+3gt} \hspace{.5cm}
s_y=\frac{\sqrt{(g+2t)(3gh^2+4g^2t-2h^2t)}}{g^2+h^2+3gt} \hspace{.5cm}
s_z=-\frac{g^2-h^2+gt}{g^2+h^2+3gt}.
\end{equation}
Inserting Eq.(\ref{s-0-3}) into Eq.(\ref{eigenvalue-1}) gives the eigenvalue for this case
as follows:
\begin{equation}
\label{ev-0-4}
\mu_2^2 = \frac{g(gh^2+4t^3)}{g^2+h^2+3gt}.
\end{equation}
It is easy to show that the choice of other sign in $\sin\theta$ and $\sin\varphi$ does not
change the eigenvalue $\mu_2^2$.

The eigenvalues for $\gamma=0$ case are summarized in Table I.

\begin{center}
\begin{tabular}{c||c|c|c} \hline
name & eigenvalue & $\lambda$ &  available region  \\  \hline \hline
$\mu_P^2$  &  $g^2$  & $2 (g^2 - t^2)$  & all     \\    \hline
$\mu_1^2$  &  $\frac{g^2 h^2 + t^2 (g + t)^2}{h^2 + (g + t)^2}$  &  $0$ & all   \\  \hline
$\mu_+^2$  & $\frac{(h r_+ + 4 t^2)^2}{r_+^2 + 4 t^2}$  &  $h r_+ + 2 t (g + t)$  &
$g \leq 2 t + h^2/(4t)$                                     \\   \hline
$\mu_-^2$  &  $\frac{(h r_- + 4 t^2)^2}{r_-^2 + 4 t^2}$  &  $h r_- + 2 t (g + t)$  &
$g \leq 2 t + h^2/(4t)$                                     \\    \hline
$\mu_2^2$  &  $\frac{g (g h^2 + 4 t^3)}{g^2 + h^2 + 3 g t}$  &  $2 t (t - g)$  &
$(3 g - 2 t) h^2 + 4 g^2 t \geq 0$                         \\   \hline
\end{tabular}

\vspace{0.1cm}
Table I: Eigenvalues for $\gamma = 0$ case
\end{center}
\vspace{0.5cm}

\subsection{$\gamma = \pi/2$ case}
For this case Eq.(\ref{main-2}) reduces to
\begin{subequations}
\label{ga-pi-2}
\begin{equation}
\label{ga-pi-2x}
2 t(g+t)\sin\theta\cos\varphi = \lambda\sin\theta\cos\varphi,
\end{equation}
\begin{equation}
\label{ga-pi-2y}
-2t(g-t)\sin\theta\sin\varphi + 2ht(1-\cos\theta)=\lambda\sin\theta\sin\varphi,
\end{equation}
\begin{equation}
\label{ga-pi-2z}
(g^2-t^2)(1+\cos\theta)-h^2(1-\cos\theta) - 2ht\sin\theta\sin\varphi =\lambda\cos\theta.
\end{equation}
\end{subequations}
Eq.(\ref{ga-pi-2x}) guarantees that the solutions for this case are categorized by
$\theta=0$, $\varphi=\pi/2$, $\varphi=3\pi/2$ and $\lambda=2t (g+t)$.
Since the calculation procedure for the first three cases are similar to the $\gamma=0$ case,
we will briefly sketch the final result only. Although the calculation procedure for the last
case is also similar to the previous case, it gives a non-trivial available region, which
is important to compute the geometric measures in next section. Therefore, we will present the
last case in detail.

When $\theta=0$, the Lagrangian multiplier constant is same with Eq.(\ref{lam-1}) and the
corresponding eigenvalue is
\begin{equation}
\label{ev-1-1}
\nu_P^2 = g^2.
\end{equation}
When $\varphi=\pi/2$, there are three types of solutions depending on $z=\tan (\theta/2)$. If
$z=(g-t)/h$, we have vanishing Lagrange multiplier constant and the corresponding eigenvalue
is
\begin{equation}
\label{ev-1-2}
\nu_1^2 = \frac{g^2h^2 + t^2(g-t)^2}{h^2+(g-t)^2}.
\end{equation}
When $z=s_{\pm} / 2t$, where
\begin{equation}
\label{til-rpm}
s_{\pm} = h \pm \sqrt{h^2 + 4t(2t + g)},
\end{equation}
the corresponding Lagrange multiplier constants are $h s_{\pm} -2t(g-t)$, and the
corresponding eigenvalues are
\begin{equation}
\label{ev-1-3}
\nu_{\pm}^2 = \frac{(h s_{\pm} + 4t^2)^2}{s_{\pm}^2 + 4t^2}.
\end{equation}
It should be noted that $\nu_{\pm}^2$ are available in entire parameter space, while
$\mu_{\pm}^2$ in $\gamma=0$ case is restricted by Eq.(\ref{avail-0-1}).
As in the case of $\gamma = 0$, $\varphi = 3\pi/2$ case does not give a new eigenvalue. This
case just reproduces $\nu_1^2$ and $\nu_{\pm}^2$.

Finally, let us discuss $\lambda = 2t(g+t)$ case. For this case Eq.(\ref{ga-pi-2x}) is
automatically solved and the remaining equations are
\begin{subequations}
\label{couple-4}
\begin{equation}
\label{couple-4y}
2ht(1 - \cos\theta) -2t(g-t)\sin\theta\sin\varphi = 2t(g+t)\sin\theta\sin\varphi
\end{equation}
\begin{equation}
\label{couple-4z}
(g^2-t^2)(1+\cos\theta)-h^2(1-\cos\theta)-2ht\sin\theta\sin\varphi=2t(g+t)\cos\theta.
\end{equation}
\end{subequations}
Since Eq.(\ref{couple-4y}) gives a relation
\begin{equation}
\label{phi-t-2}
\sin\varphi = \frac{h}{2g} \frac{1-\cos\theta}{\sin\theta},
\end{equation}
combining Eq.(\ref{couple-4z}) and Eq.(\ref{phi-t-2}) yields
\begin{equation}
\label{pi-cos-1}
\cos\theta = -\frac{g^2-h^2-gt}{g^2+h^2-3gt}.
\end{equation}
The requirement $-1\leq \cos\theta \leq 1$ gives first available condition
\begin{equation}
\label{avail-pi-1}
(g-2t) (h^2-gt) \geq 0.
\end{equation}

Now we choose $\sin\theta$ as
\begin{equation}
\label{pi-sin-1}
\sin\theta = \frac{\sqrt{4g(g-2t)(h^2-gt)}}{g^2+h^2-3gt}.
\end{equation}
Then from Eq.(\ref{phi-t-2}) $\sin\varphi$ becomes
\begin{equation}
\label{pi-phi-1}
\sin\varphi=\frac{h}{2} \sqrt{\frac{g-2t}{g(h^2-gt)}}.
\end{equation}
Another requirement $-1\leq \sin\varphi\leq 1$ gives second available condition
\begin{equation}
\label{avail-pi-2}
(g-2t) (3gh^2 - 4 g^2 t + 2 h^2 t) \geq 0.
\end{equation}
Choosing $\cos\varphi$ as
\begin{equation}
\label{pi-phi-2}
\cos\varphi = \sqrt{\frac{3gh^2-4g^2t+2h^2t}{4g (h^2-gt)}},
\end{equation}
it is straightforward to show that the eigenvalues for this case is
\begin{equation}
\label{ev-1-4}
\nu_2^2 = \frac{g(gh^2-4t^3)}{g^2+h^2-3gt}.
\end{equation}
It is easy to show that the different choices in the sign of $\sin\theta$ and/or $\cos\varphi$
do not change the eigenvalue. Although the available region for $\nu_2^2$ is restricted by
Eq.(\ref{avail-pi-1}) and Eq.(\ref{avail-pi-2}), one can show that Eq.(\ref{avail-pi-2})
implies Eq.(\ref{avail-pi-1}) already. To show this explicitly let us consider
$g \geq 2t$ case first. In this case Eq.(\ref{avail-pi-2}) imposes $h^2\geq 4g^2t/(3g+2t)$.
Therefore
$$h^2-gt \geq \frac{4g^2t}{3g+2t} - gt = \frac{gt}{3g+2t} (g-2t) \geq 0. $$
Similarly, one can show that Eq.(\ref{avail-pi-2}) implies Eq.(\ref{avail-pi-1}) for
$g \leq 2t$ region too. Therefore, the available region for $\nu_2^2$ is restricted by
Eq.(\ref{avail-pi-2}) only.

The eigenvalues in $\gamma=\pi/2$ case is summarized in
Table II.

\begin{center}
\begin{tabular}{c||c|c|c} \hline
name  &  eigenvalue & $\lambda$ &  available region  \\  \hline \hline
$\nu_P^2$  &  $g^2$  & $2 (g^2 - t^2)$ &  all     \\    \hline
$\nu_1^2$  &  $\frac{g^2 h^2 + t^2 (g - t)^2}{h^2 + (g - t)^2}$  &  $0$  & all
                                              \\    \hline
$\nu_+^2$  &  $\frac{(h s_+ + 4 t^2)^2}{s_+^2 + 4 t^2}$  &
$h s_+ - 2 t (g - t)$  & all
                                                \\     \hline
$\nu_-^2$  &  $\frac{(h s_- + 4 t^2)^2}{s_-^2 + 4 t^2}$  &
$h s_- - 2 t (g - t)$  & all
                                                \\     \hline
$\nu_2^2$  &  $\frac{g (g h^2 - 4 t^3)}{g^2 + h^2 - 3 g t}$  &  $2 t (g+t)$
& $(g-2t)(3gh^2-4g^2t+2h^2t) \geq 0$
                                                              \\   \hline
\end{tabular}

\vspace{0.1cm}
Table II: Eigenvalues for $\gamma = \pi / 2$ case
\end{center}
\vspace{0.5cm}

\subsection{$h\rightarrow 0$ limit}
Since $|\psi\ra$ is independent of $\gamma$ in the $h \rightarrow 0$ limit, all eigenvalues
for $\gamma = 0$ and $\gamma = \pi/2$ cases should be same including the available
region in the parameter space. Note that $\mu_+^2 = \mu_-^2$ and
$\nu_+^2 = \nu_-^2$ in the $h \rightarrow 0$ limit. In this limit the
eigenvalues for $\gamma=0$ exactly coincide with eigenvalues for $\gamma=\pi/2$ as
following:
\begin{equation}
\label{0h0}
\mu_P^2=\nu_P^2=g^2   \hspace{.5cm}
\mu_1^2=\nu_1^2 = t^2  \hspace{.5cm}
\mu_2^2=\nu_{\pm}^2=\frac{4t^3}{3t+g}  \hspace{.5cm}
\mu_{\pm}^2=\nu_2^2=\frac{4t^3}{3t-g}.
\end{equation}
In addition, first three eigenvalues in Eq.(\ref{0h0}) are available in the full parameter
space and the last one is available only at $g \leq 2 t$. Thus, our calculational results are
perfectly consistent in the $h \rightarrow 0$ limit.

\section{Geometric Measure for $\gamma = 0$}

\begin{figure}[ht!]
\begin{center}
\includegraphics[height=6cm]{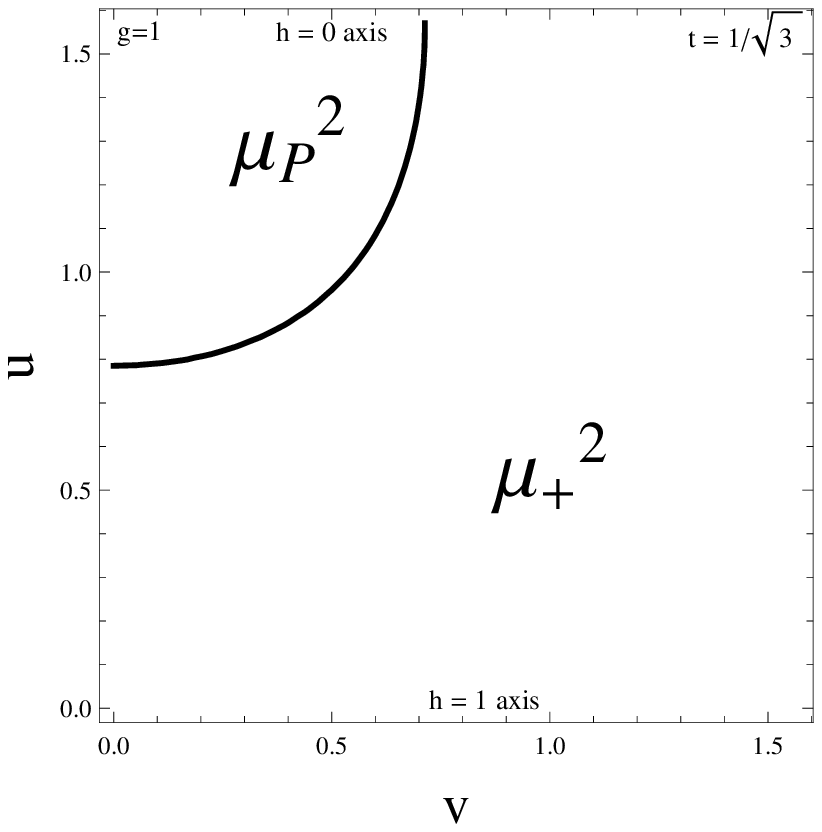}
\includegraphics[height=6cm]{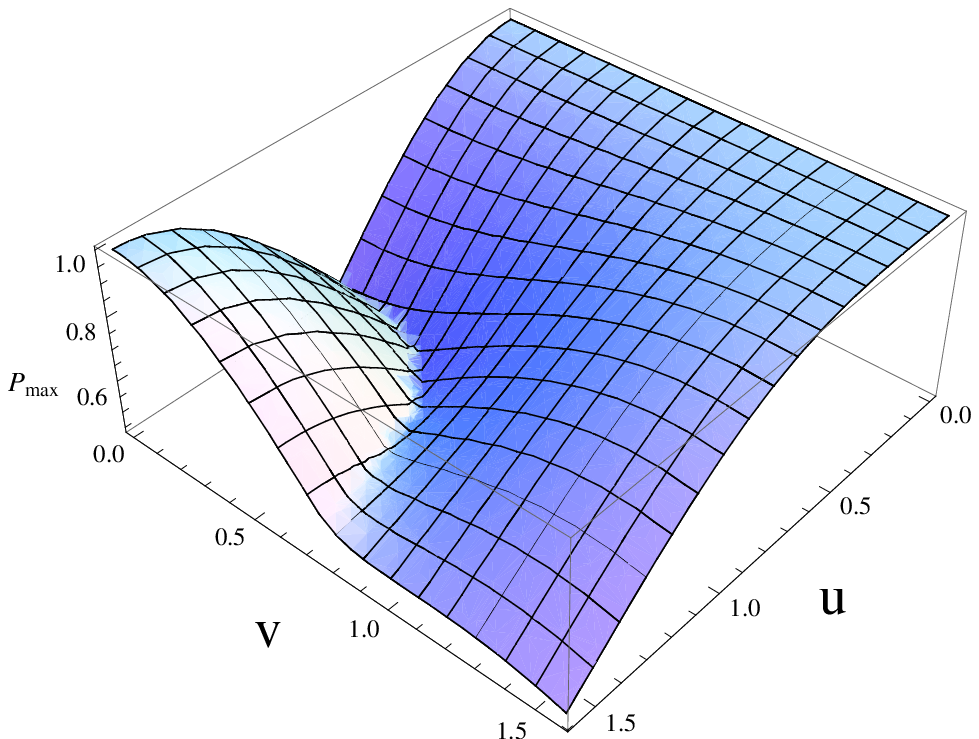}
\includegraphics[height=6cm]{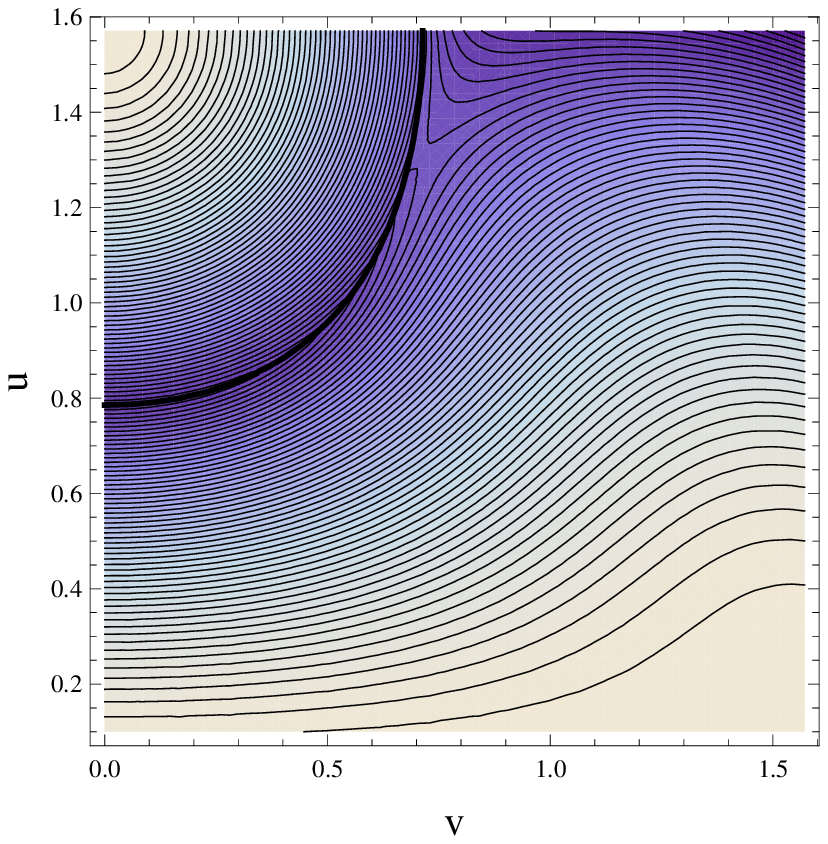}
\caption[fig1]{(Color online)
Fig. 1a is a plot of the applicable domains in $(u,v)$-plane for $\gamma = 0$. 
The principal domain 
$P_{max}=\mu_P^2$ is located in small $v$ and large $u$ region. This fact indicates that
this domain is around large $g$ region. Fig. 1b is plot of $(u,v)$-dependence of $P_{max}$
for $\gamma = 0$ case. Many highly entangled states are represented as a valley in this
figure. Around $u \sim 0$ and $(u \sim \pi/2, v \sim 0)$ there are a lot of less entangled
states. To compare the applicable domains with $P_{max}$ we plot both simultaneously in the
$(u,v)$ plane in Fig. 1c. The black thick line is a boundary between domains. The blue-color
and white-color represent the highly- and less-entangled states respectively. Fig. 1c shows
that the highly-entangled states reside around the boundary between domains.}
\end{center}
\end{figure}

In this section we would like to compute the geometric entanglement measure defined
\begin{equation}
\label{4-geo-1}
G(\psi) = 1 - P_{max} (\psi)
\end{equation}
for $\gamma = 0$ case. In order to compute $P_{max}$ we would like to emphasize three points,
which simplify the following calculation. Firstly, note that $P_{max}$ is given by
\begin{equation}
\label{4-pmax-1}
P_{max} = \max (\mu_i^2).
\end{equation}
Therefore, we should choose the largest eigenvalue from all eigenvalues, each of which has
its own available regions in the parameter space. Secondly, note that
\begin{equation}
\label{4-eigen-1}
\mu_+^2 - \mu_-^2 = \frac{128 h t^{7/2}}{(r_+^2 + 4 t^2) (r_-^2 + 4 t^2)}
\left(2 t + \frac{h^2}{4 t} - g \right)^{3/2}.
\end{equation}
This means that $\mu_-^2$ is always smaller than $\mu_+^2$ in the available region
$g \leq 2t+h^2/(4t)$. Therefore, we can exclude $\mu_-^2$ from beginning for the computation
of $P_{max}$. Thirdly, note that $P_{max}$ is obtained from the eigenvalues whose Lagrange
multiplier constants are positive\cite{analytic}.
This fact excludes $\mu_1^2$ too. Considering all of these
facts and available regions, it is convenient to divide the whole parameter space into the
following four regions:
\begin{eqnarray}
\label{4-region-1}
& & (\mbox{region I})\hspace{.5cm} g \geq 2 t + \frac{h^2}{4 t}:
                                                          \hspace{.5cm}P_{max} = \mu_P^2
                                                       \\   \nonumber
& & (\mbox{region II})\hspace{.5cm} t \leq g \leq 2t + \frac{h^2}{4t}:
\hspace{.5cm} P_{max} = \max(\mu_P^2, \mu_+^2)
                                                               \\   \nonumber
& & (\mbox{region III})\hspace{.5cm} g \leq t \hspace{.2cm} \& \hspace{.2cm} {\cal C}_1 \geq 0:
\hspace{.5cm}   P_{max} = \max(\mu_+^2, \mu_2^2)
                                                               \\    \nonumber
& & (\mbox{region IV})\hspace{.5cm} g \leq t \hspace{.2cm} \& \hspace{.2cm} {\cal C}_1 \leq 0:
\hspace{.5cm} P_{max} = \mu_+^2
\end{eqnarray}
where
\begin{equation}
\label{4-cri-1}
{\cal C}_1 = (3 g - 2t)h^2 + 4 g^2 t.
\end{equation}

In order to compare $\mu_+^2$ with $\mu_2^2$ we compute $\mu_+^2 - \mu_2^2$, which is
\begin{equation}
\label{4-eigen-2}
\mu_+^2 - \mu_2^2 = \frac{2}{(r_+^2 + 4t^2) (g^2+h^2+3gt)} \left(\alpha_1 + \beta_1
\sqrt{h^2 + 4t(2t-g)} \right)
\end{equation}
where
\begin{eqnarray}
\label{4-eigen-3}
& &\alpha_1 = h^6+gh^4t+8h^4t^2+20gh^2t^3+16g^2t^4+4h^2t^2(2t^2-g^2)
                                                             \\   \nonumber
& &\beta_1 = h(h^4+3gh^2t+4g^2t^2+4h^2t^2+8gt^3).
\end{eqnarray}
Since the last term in $\alpha_1$, $4h^2t^2(2t^2-g^2)$, is non-negative in the region
$g\leq t$, both $\alpha_1$ and $\beta_1$ are non-negative in region III. In region III,
therefore, $P_{max}$ becomes $\mu_+^2$.

In region II it has been shown in Ref.\cite{maximal} that $\mu_P^2 = \mu_+^2$ when
${\cal D}_1 = 0$, where
\begin{equation}
\label{4-cri-2}
{\cal D}_1 = gh^2 - (g+t)^2 (g-2t).
\end{equation}
Therefore, the region II should be divided into two regions, {\it i.e.} ${\cal D}_1 \geq 0$
and ${\cal D}_1 \leq 0$. Simple consideration shows that $\mu_P^2 \geq \mu_+^2$ when
${\cal D}_1 \leq 0$ and $\mu_P^2 \leq \mu_+^2$ when ${\cal D}_1 \geq 0$. Combining all of
these facts, one can conclude
\begin{eqnarray}
\label{4-region-2}
& &(\mbox{region A})\hspace{.5cm} g\geq 2t+\frac{h^2}{4t}:\hspace{.5cm}P_{max}=\mu_P^2
                                                          \\   \nonumber
& &(\mbox{region B})\hspace{.5cm}t \leq g \leq 2t+\frac{h^2}{4t} \hspace{.2cm} \& \hspace{.2cm}
{\cal D}_1 \leq 0: \hspace{.5cm} P_{max}=\mu_P^2
                                                          \\   \nonumber
& &(\mbox{region C}) \hspace{.5cm}t \leq g \leq 2t+\frac{h^2}{4t} \hspace{.2cm} \& \hspace{.2cm}
{\cal D}_1 \geq 0: \hspace{.5cm} P_{max} = \mu_+^2                      \\   \nonumber
& &(\mbox{region D}) \hspace{.5cm} g \leq t: \hspace{.5cm} P_{max} = \mu_+^2.
\end{eqnarray}

Now, we would like to unify the regions as many as possible to simplify the expression of
$P_{max}$. First, one can show that ${\cal D}_1$ is always non-positive in region A as
following. Since $h^2 \leq 4t(g-2t)$ in region A, in this region
\begin{equation}
\label{4-cri-3}
{\cal D}_1 = gh^2 - (g+t)^2 (g-2t) \leq -(g-2t) (g-t)^2 \leq 0.
\end{equation}
Second, one can show easily that ${\cal D}_1$ is always non-negative at region D as following.
In this region
\begin{equation}
\label{4-cri-4}
{\cal D}_1 = gh^2 + (g+t)^2 (2t-g) \geq 0
\end{equation}
because both terms are non-negative. Combining these facts and Eq.(\ref{4-region-2}) makes
$P_{max}$ to be expressed as
\begin{eqnarray}
\label{4-final}
P_{max} = \left\{     \begin{array}{cc}
                \mu_P^2  &  \hspace{.5cm} \mbox{when  } {\cal D}_1 \leq 0    \\
               \mu_+^2   &  \hspace{.5cm} \mbox{when  } {\cal D}_1 \geq 0.
                      \end{array}                                  \right.
\end{eqnarray}

In order to understand the behavior of $P_{max}$ more clearly we introduce the two
parameters $u$ and $v$ as following:
\begin{equation}
\label{4-para-1}
g = \sin u \cos v, \hspace{.7cm} t = \sin u \sin v / \sqrt{3} \hspace{.7cm}
h = \cos u
\end{equation}
with $0 \leq u, v \leq \pi/2$. Then, one can plot the applicable domains ${\cal D}_1 \leq 0$
and ${\cal D}_1 \geq 0$ in the $u-v$ plane, which is Fig. 1a. As Fig. 1a has shown, the domain
for ${\cal D}_1 \leq 0$ is biased in the small $v$ and large $u$ region. This indicates that
the domains for ${\cal D}_1 \leq 0$ is around large $g$ region. The remaining region is
the domain for ${\cal D}_1 \geq 0$. As will be shown in next section, the number of the 
applicable domains for $\gamma = \pi/2$ case is not two but three. This means that the phase
factor $\gamma$ has great impact in the geometric measure of entanglement.

Fig. 1b is $(u,v)$-dependence of $P_{max}$ given in Eq.(\ref{4-final}). At $u=0$, which means
$h=1$, $P_{max}$ becomes $1$ because it is separable state. At $v=0$ and $u=\pi/2$, which means
that $g=1$, $P_{max}$ becomes $1$ again. Between them there is valley, which represents the
set of the highly entangled states. There is different kind of the highly entangled states
around $u=v=\pi/2$. These highly entangled states are states located near W-state,
$|W\ra = (1/\sqrt{3})(|011\ra + |101\ra + |110\ra)$.

In order to compare $P_{max}$ with the applicable domains we plot $P_{max}$ and the boundary
of domains simultaneously in $u-v$ plane in Fig. 1c. In Fig. 1c the black thick line is a
boundary of the domains. The thick-blue color and light-blue (or white) colors represent the
highly-entangled and less-entangled states, respectively. In the right-upper corner there are
many highly entangled states which are located near W-state. Another type of the
highly entangled states reside near the boundary of the applicable domains. Apart from the
boundary more and more the quantum states lose the entanglement, and eventually reduce to
the separable state.

Now, we consider several special cases. First example is $t=1/\sqrt{3}$ and $g=h=0$. In this
case ${\cal D}_1=2\sqrt{3}/9 > 0$ and $r_+=\sqrt{8/3}$,
which gives $P_{max}=4/9$. Second example is $t=0$ and $g \geq h$. In this case
${\cal D}_1 = -g(g^2-h^2) \leq 0$ and $P_{max} = g^2$. Third example is $t=0$ and
$g \leq h$. In this case ${\cal D}_1 = g(h^2-g^2) \geq 0$ and $r_+=2h$, which gives
$P_{max}=h^2$. The second and third examples are consistent with
$P_{max}(GHZ) = \max(|\alpha|^2, |\beta|^2)$, where $|GHZ\ra=\alpha |000\ra + \beta |111\ra$.
Fourth example is $g=0$ case. In this case ${\cal D}_1 = 2t^3 \geq 0$ and
$r_+ = h + \sqrt{h^2 + 8t^2}$, which results in
\begin{equation}
\label{4-ex-1}
P_{max} = \frac{(h^4+8h^2t^2+8t^4) + h(h^2+4t^2) \sqrt{h^2+8t^2}}
{(h^2+6t^2) + h\sqrt{h^2+8t^2}}.
\end{equation}
One can show that various limits of Eq.(\ref{4-ex-1}) are consistent with the previously
derived results. The last example is $h=0$ case. In this case it is easy to show
\begin{eqnarray}
\label{4-ex-2}
P_{max} = \left\{            \begin{array}{cc}
                  g^2  & \hspace{.5cm} \mbox{when   }  g \geq 2t  \\
                4t^3/(3t - g)  &  \hspace{.5cm} \mbox{when   } g \leq 2t.
                             \end{array}                  \right.
\end{eqnarray}
Eq.(\ref{4-ex-2}) is perfectly in agreement with the result of Ref.\cite{shared}.

\section{Geometric Measure for $\gamma = \pi/2$}

In this section we would like to compute the geometric entanglement measure for
$\gamma = \pi/2$ case. From the constraint of the positive Lagrange multiplier constant
we can exclude $\nu_1^2$ and $\nu_-^2$ from beginning stage for the computation of the
geometric measure. Next, we should examine the sign of the Lagrange multiplier
constant for $\nu_+^2$, that is
\begin{equation}
\label{5-Lagrange-1}
\lambda_+ = h s_+ -2t(g-t).
\end{equation}
It is easy to show that $\lambda_+ \geq 0$ in $g \leq t$ region. Also it is straightforward
to show that $\lambda_+ \geq 0$ when ${\cal C}_+ \geq 0$ and $\lambda_+ \leq 0$ when
${\cal C}_+ \leq 0$, where
\begin{equation}
\label{5-cri-1}
{\cal C}_+ = h^2 (2g+t) - t(g-t)^2.
\end{equation}
Examining Table II and Eq.(\ref{5-cri-1}) leads us to divide the whole parameter space into
the following ten regions:
\begin{eqnarray}
\label{5-region-1}
& & \hspace{3.0cm} (i) \hspace{.3cm} g \geq 2 t              \\   \nonumber
& & (\mbox{region I})\hspace{.5cm} {\cal C}_2 \leq 0 \hspace{.2cm} \& \hspace{.2cm}
{\cal C}_+ \leq 0:   \hspace{.5cm} P_{max} = \nu_P^2
                                                       \\   \nonumber
& & (\mbox{region II})\hspace{.5cm} {\cal C}_2 \geq 0 \hspace{.2cm} \& \hspace{.2cm}
{\cal C}_+ \leq 0:   \hspace{.5cm} P_{max} = \max (\nu_P^2, \nu_2^2)
                                                       \\   \nonumber
& & (\mbox{region III})\hspace{.5cm} {\cal C}_2 \leq 0 \hspace{.2cm} \& \hspace{.2cm}
{\cal C}_+ \geq 0:   \hspace{.5cm} P_{max} = \max (\nu_P^2, \nu_+^2)
                                                       \\   \nonumber
& & (\mbox{region IV})\hspace{.5cm} {\cal C}_2 \geq 0 \hspace{.2cm} \& \hspace{.2cm}
{\cal C}_+ \geq 0:   \hspace{.5cm} P_{max} = \max (\nu_P^2, \nu_+^2, \nu_2^2)
                                                       \\   \nonumber
& & \hspace{3.0cm} (ii) \hspace{.3cm} t \leq g \leq 2 t              \\   \nonumber
& & (\mbox{region V})\hspace{.5cm} {\cal C}_2 \geq 0 \hspace{.2cm} \& \hspace{.2cm}
{\cal C}_+ \leq 0:   \hspace{.5cm} P_{max} = \nu_P^2
                                                       \\   \nonumber
& & (\mbox{region VI})\hspace{.5cm} {\cal C}_2 \leq 0 \hspace{.2cm} \& \hspace{.2cm}
{\cal C}_+ \leq 0:   \hspace{.5cm} P_{max} = \max (\nu_P^2, \nu_2^2)
                                                       \\   \nonumber
& & (\mbox{region VII})\hspace{.5cm} {\cal C}_2 \geq 0 \hspace{.2cm} \& \hspace{.2cm}
{\cal C}_+ \geq 0:   \hspace{.5cm} P_{max} = \max (\nu_P^2, \nu_+^2)
                                                       \\   \nonumber
& & (\mbox{region VIII})\hspace{.5cm} {\cal C}_2 \leq 0 \hspace{.2cm} \& \hspace{.2cm}
{\cal C}_+ \geq 0:   \hspace{.5cm} P_{max} = \max (\nu_P^2, \nu_+^2, \nu_2^2)
                                                       \\   \nonumber
& & \hspace{3.0cm} (iii) \hspace{.3cm} g \leq  t              \\   \nonumber
& & (\mbox{region IX})\hspace{.5cm} {\cal C}_2 \leq 0: P_{max} = \max (\nu_+^2, \nu_2^2)
                                                       \\   \nonumber
& & (\mbox{region X})\hspace{.5cm} {\cal C}_2 \geq 0: P_{max} = \nu_+^2
\end{eqnarray}
where
\begin{equation}
\label{5-cri-2}
{\cal C}_2 = (3g+2t) h^2 - 4 g^2 t.
\end{equation}

\begin{center}
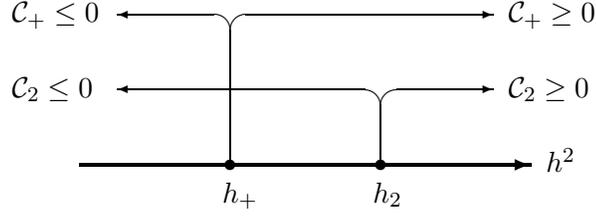
\begin{figure}
\setlength{\unitlength}{1cm}
\begin{picture}(7,3)
        \put(3.0, 0.5){\line(0,1){1.75}}
	\put(2.5, 2.25){\oval(1.0, 0.5)[tr]}
	\put(3.5, 2.25){\oval(1.0, 0.5)[tl]}
	\put(2.5, 2.5){\vector(-1,0){1.0}}
	\put(3.5, 2.5){\vector(1,0){3.0}}
	
        \put(5.0, 0.5){\line(0,1){0.75}}
	\put(4.5, 1.25){\oval(1.0, 0.5)[tr]}
	\put(5.5, 1.25){\oval(1.0,0.5)[tl]}
	\put(4.5, 1.5){\vector(-1, 0){3.0}}
	\put(5.5, 1.5){\vector( 1, 0){1.0}}
	
	\put(3.0, 0.5){\circle*{0.15}}
	\put(5.0, 0.5){\circle*{0.15}}
	\thicklines
	\put(1.0, 0.5){\vector(1,0){6}}
	
	\put(2.9, 0.0){$h_+$}
	\put(4.9, 0.0){$h_2$}
	\put(7.2, 0.4){$h^2$}
	
	\put(6.7, 1.4){${\cal C}_2 \geq 0$}
	\put(6.7, 2.4){${\cal C}_+ \geq 0$}
	
	\put(0.1, 1.4){${\cal C}_2 \leq 0$}
	\put(0.1, 2.4){${\cal C}_+ \leq 0$}
\end{picture}
\caption[fig2]{Pictorial representation for ${\cal C}_2 \geq 0$, ${\cal C}_2 \leq 0$,
${\cal C}_+ \geq 0$, and ${\cal C}_+ \leq 0$ when $g \geq t$.}
\end{figure}
\end{center}
Although the whole space is divided into the ten regions, one can show that some regions do not
exist. In order to show this it is convenient to introduce
\begin{equation}
\label{5-intro-1}
h_2 = \frac{4 g^2 t}{3g+2t}   \hspace{2.0cm}  h_+ = \frac{t (g-t)^2}{2g+t}.
\end{equation}
Then, their difference becomes
\begin{equation}
\label{5-intro-2}
h_2 - h_+ = \frac{t (g+t)^2}{(3g+2t) (2g+t)} (5g-2t).
\end{equation}
Eq.(\ref{5-intro-2}) implies that $h_2 \geq h_+$ in the region $g\geq t$. Then the regions
${\cal C}_2 \geq 0$, ${\cal C}_2 \leq 0$, ${\cal C}_+ \geq 0$, and ${\cal C}_+ \leq 0$
when $g\geq t$ can be represented as Fig. 2. With an help of Fig. 2 it is easy to
understand that there is no region which satisfies both ${\cal C}_2 \geq 0$ and
${\cal C}_+ \leq 0$ when $g \geq t$. This implies that region II and region V do not exist
in the whole parameter space.

In order to compare $\nu_P^2$ with $\nu_2^2$ we compute $\nu_P^2 - \nu_2^2$, which is
\begin{equation}
\label{5-eigen-1}
\nu_P^2 - \nu_2^2 = \frac{g (g+t) (g-2t)^2}{g^2+h^2-3gt}.
\end{equation}
Therefore, the sign of $\nu_P^2 - \nu_2^2$ is determined by $g^2+h^2-3gt$. If
${\cal C}_2 \geq 0$, $h^2 \geq h_2$ and
\begin{equation}
\label{5-eigen-2}
g^2+h^2-3gt \geq \frac{3g (g-2t) (g+t)}{3g+2t}.
\end{equation}
Therefore, if ${\cal C}_2 \geq 0$ in $g \geq 2t$ region, $\nu_P^2 \geq \nu_2^2$. Thus, we
can exclude $\nu_2^2$ in region IV. Similarly, one can show that if ${\cal C}_2 \leq 0$ in
$t \leq g \leq 2t$ region, $\nu_P^2 \leq \nu_2^2$. Therefore, we can exclude $\nu_P^2$
in regions VI and VIII.
\begin{center}
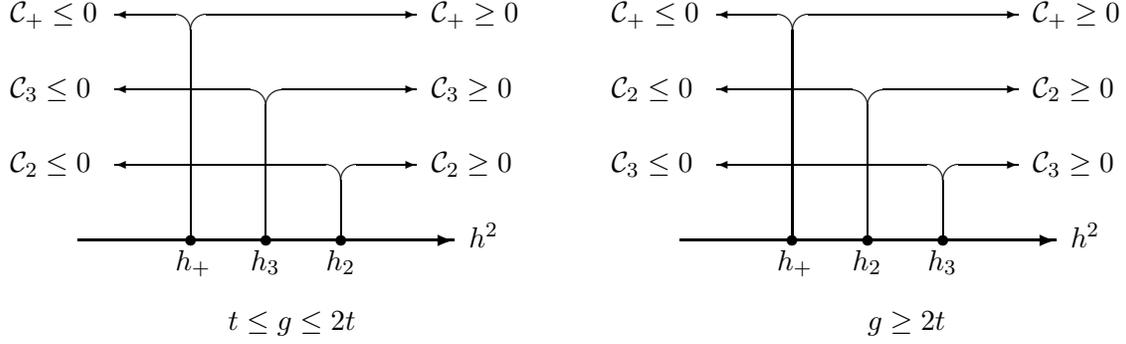
\begin{figure}
\setlength{\unitlength}{1cm}
\begin{picture}(15,5)
        \put(2.5, 1.5){\line(0,1){2.75}}
	\put(2.0, 4.25){\oval(1.0, 0.5)[tr]}
	\put(3.0, 4.25){\oval(1.0, 0.5)[tl]}
	\put(2.0, 4.5){\vector(-1,0){0.5}}
        \put(3.0, 4.5){\vector( 1,0){2.5}}

        \put(3.5, 1.5){\line(0,1){1.75}}
	\put(3.0, 3.25){\oval(1.0, 0.5)[tr]}
	\put(4.0, 3.25){\oval(1.0, 0.5)[tl]}
	\put(3.0, 3.5){\vector(-1,0){1.5}}
	\put(4.0, 3.5){\vector( 1,0){1.5}}
	
        \put(4.5, 1.5){\line(0,1){0.75}}
	\put(4.0, 2.25){\oval(1.0, 0.5)[tr]}
	\put(5.0, 2.25){\oval(1.0,0.5)[tl]}
	\put(4.0, 2.5){\vector(-1, 0){2.5}}
	\put(5.0, 2.5){\vector( 1,0){0.5}}
	
	\put(2.5, 1.5){\circle*{0.15}}
	\put(3.5, 1.5){\circle*{0.15}}
	\put(4.5, 1.5){\circle*{0.15}}
	\thicklines
	\put(1.0, 1.5){\vector(1,0){5}}
	
	\put(2.3, 1.1){$h_+$}
	\put(3.3, 1.1){$h_3$}
	\put(4.3, 1.1){$h_2$}
	\put(6.2, 1.4){$h^2$}
	
	\put(5.7, 2.4){${\cal C}_2 \geq 0$}
	\put(5.7, 3.4){${\cal C}_3 \geq 0$}
	\put(5.7, 4.4){${\cal C}_+ \geq 0$}
	
	\put(0.1, 2.4){${\cal C}_2 \leq 0$}
	\put(0.1, 3.4){${\cal C}_3 \leq 0$}
	\put(0.1, 4.4){${\cal C}_+ \leq 0$}
	
	\put(3.0, 0.3){$t \leq g \leq 2 t$}
        \thinlines
        \put(10.5, 1.5){\line(0,1){2.75}}
	\put(10.0, 4.25){\oval(1.0, 0.5)[tr]}
	\put(11.0, 4.25){\oval(1.0, 0.5)[tl]}
	\put(10.0, 4.5){\vector(-1,0){0.5}}
        \put(11.0, 4.5){\vector( 1,0){2.5}}

        \put(11.5, 1.5){\line(0,1){1.75}}
	\put(11.0, 3.25){\oval(1.0, 0.5)[tr]}
	\put(12.0, 3.25){\oval(1.0, 0.5)[tl]}
	\put(11.0, 3.5){\vector(-1,0){1.5}}
	\put(12.0, 3.5){\vector( 1,0){1.5}}
	
        \put(12.5, 1.5){\line(0,1){0.75}}
	\put(12.0, 2.25){\oval(1.0, 0.5)[tr]}
	\put(13.0, 2.25){\oval(1.0,0.5)[tl]}
	\put(12.0, 2.5){\vector(-1, 0){2.5}}
	\put(13.0, 2.5){\vector( 1,0){0.5}}
	
	\put(10.5, 1.5){\circle*{0.15}}
	\put(11.5, 1.5){\circle*{0.15}}
	\put(12.5, 1.5){\circle*{0.15}}
	\thicklines
	\put(9.0, 1.5){\vector(1,0){5}}
	
	\put(10.3, 1.1){$h_+$}
	\put(11.3, 1.1){$h_2$}
	\put(12.3, 1.1){$h_3$}
	\put(14.2, 1.4){$h^2$}
	
	\put(13.7, 2.4){${\cal C}_3 \geq 0$}
	\put(13.7, 3.4){${\cal C}_2 \geq 0$}
	\put(13.7, 4.4){${\cal C}_+ \geq 0$}
	
	\put(8.1, 2.4){${\cal C}_3 \leq 0$}
	\put(8.1, 3.4){${\cal C}_2 \leq 0$}
	\put(8.1, 4.4){${\cal C}_+ \leq 0$}
	
	\put(11.5, 0.3){$g \geq 2 t$}
\end{picture}
\caption[fig3]{Pictorial representation for ${\cal C}_2 \geq 0$, ${\cal C}_2 \leq 0$,
${\cal C}_+ \geq 0$, ${\cal C}_+ \leq 0$, ${\cal C}_3 \geq 0$ and ${\cal C}_3 \leq 0$
when $t \leq g \leq 2t$ (Fig. 2 a) and $g \geq 2 t$ (Fig. 2 b).}
\end{figure}
\end{center}
Next, we compute $\nu_P^2 - \nu_+^2$, which is
\begin{equation}
\label{5-eigen-3}
\nu_P^2 - \nu_+^2 = \frac{2}{s_+^2 + 4t^2} \left(\alpha_2 + \beta_2 \sqrt{h^2+4t(2t+g)}\right)
\end{equation}
where
\begin{eqnarray}
\label{5-eigen-4}
& &\alpha_2 = -h^4+(g+2t)(g-4t)h^2+2t(g-t)(g+2t)^2    \\   \nonumber
& &\beta_2 = h (g^2-h^2-4t^2).
\end{eqnarray}
Direct calculation shows that in $g\geq t$ region $\nu_P^2 = \nu_+^2$ when ${\cal C}_3 = 0$,
where
\begin{equation}
\label{5-cri-3}
{\cal C}_3 = gh^2-(g-t)^2(g+2t).
\end{equation}
In addition, simple consideration shows that in $g \geq t$ region $\nu_P^2 \geq \nu_+^2$
when ${\cal C}_3 \leq 0$ and $\nu_P^2 \leq \nu_+^2$ when ${\cal C}_3 \geq 0$.

In order to check which eigenvalue is dominant in each region it is convenient to
introduce another parameter
\begin{equation}
\label{5-intro-3}
h_3 = \frac{(g-t)^2 (g+2t)}{g}.
\end{equation}
Then, it is easy to show
\begin{eqnarray}
\label{5-intro-4}
& & h_+ \leq h_2 \leq h_3  \hspace{2.0cm} \mbox{when} \hspace{.2cm} 2t \leq g
                                                            \\   \nonumber
& &h_+ \leq h_3 \leq h_2  \hspace{2.0cm} \mbox{when} \hspace{.2cm} t \leq g \leq 2t.
\end{eqnarray}
Eq.(\ref{5-intro-4}) enables us to represent ${\cal C}_2 \geq 0$, ${\cal C}_2 \leq 0$,
${\cal C}_+ \geq 0$, ${\cal C}_+ \leq 0$, ${\cal C}_3 \geq 0$ and ${\cal C}_3 \leq 0$
in one-dimensional coordinate, which is illustrated in Fig. 3.
With an help of Fig. 3 one can show
easily that in region III ${\cal C}_3$ is always non-positive and therefore, $P_{max}$
becomes $\nu_P^2$. Using Fig. 3a $P_{max}$ in region VII is $\nu_+^2$. Using Fig. 3b again
one can show that region IV is divided into
\begin{eqnarray}
\label{5-region-2}
& &(\mbox{region IV-a}) \hspace{.5cm} {\cal C}_2 \geq 0 \hspace{.2cm} \& \hspace{.2cm}
{\cal C}_3 \leq 0: \hspace{.5cm} P_{max} = \nu_P^2
                                                          \\   \nonumber
& &(\mbox{region IV-b}) \hspace{.5cm} {\cal C}_2 \geq 0 \hspace{.2cm} \& \hspace{.2cm}
{\cal C}_3 \geq 0: \hspace{.5cm} P_{max} = \nu_+^2.
\end{eqnarray}

Finally, we compute $\nu_+^2 - \nu_2^2$, which is
\begin{equation}
\label{5-eigen-5}
\nu_+^2 - \nu_2^2 = \frac{2}{(s_+^2 + 4t^2) (g^2+h^2-3gt)} \left( \alpha_3 + \beta_3
\sqrt{h^2+4t(2t+g)} \right)
\end{equation}
where
\begin{eqnarray}
\label{5-eigen-6}
& &\alpha_3 = h^6+t(8t-g)h^4-4t^2(g^2+5gt-2t^2)h^2+16g^2t^4     \\   \nonumber
& &\beta_3 = h \left[h^4+t(4t-3g)h^2+4gt^2(g-2t) \right].
\end{eqnarray}
One can show directly that $\nu_+^2 - \nu_2^2 = 0$ when ${\cal C}_2=0$. Also, it is
straightforward to show that in $g \leq 2t$ region $\nu_+^2$ is always smaller than $\nu_2^2$.
Therefore, we can exclude $\nu_+^2$ in regions VIII and IX. Combining all of these facts,
one can express $P_{max}$ for $\gamma = \pi/2$ case as follows:
\begin{eqnarray}
\label{5-final}
& &\hspace{3cm} (i) \hspace{.3cm} g \geq 2t                  \\   \nonumber
& & P_{max} = \left\{            \begin{array}{cc}
                     \nu_+^2 & \hspace{2.0cm} {\cal C}_2 \geq 0 \hspace{.2cm} \& \hspace{.2cm}
                                              {\cal C}_3 \geq 0    \\
                     \nu_P^2 & \hspace{2.0cm} \mbox{remaining region}
                                 \end{array}                         \right.
                                                             \\    \nonumber
& &\hspace{3cm} (i) \hspace{.3cm} g \leq 2t                  \\   \nonumber
& & P_{max} = \left\{            \begin{array}{cc}
                           \nu_+^2 & \hspace{2.0cm} {\cal C}_2 \geq 0   \\
                           \nu_2^2 &  \hspace{2.0cm} {\cal C}_2 \leq 0.
                                 \end{array}                          \right.
\end{eqnarray}

\begin{figure}[ht!]
\begin{center}
\includegraphics[height=6cm]{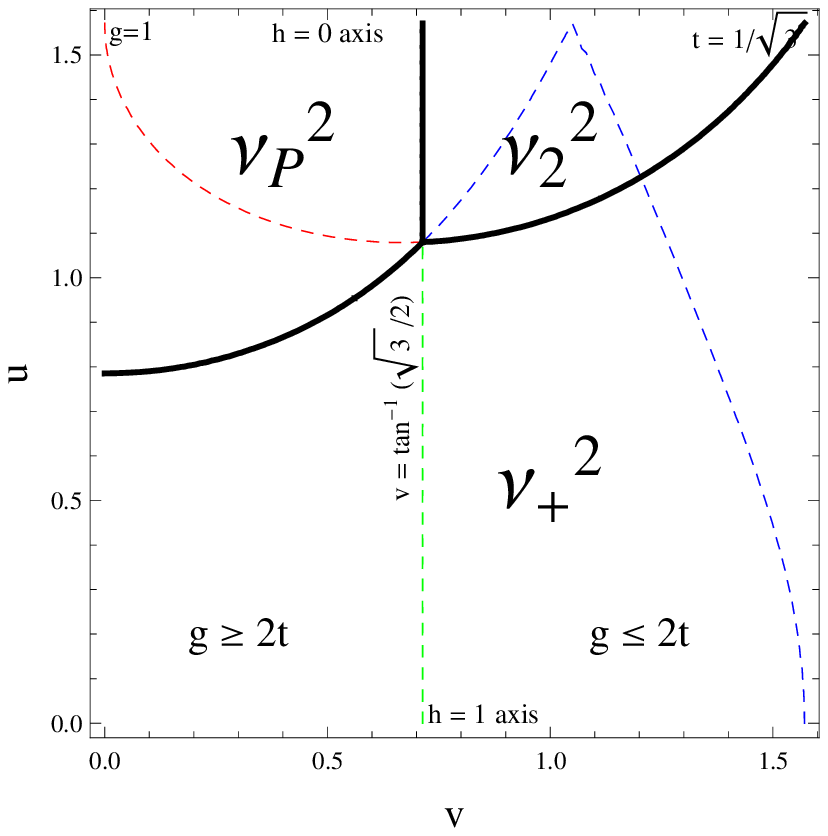}
\includegraphics[height=6cm]{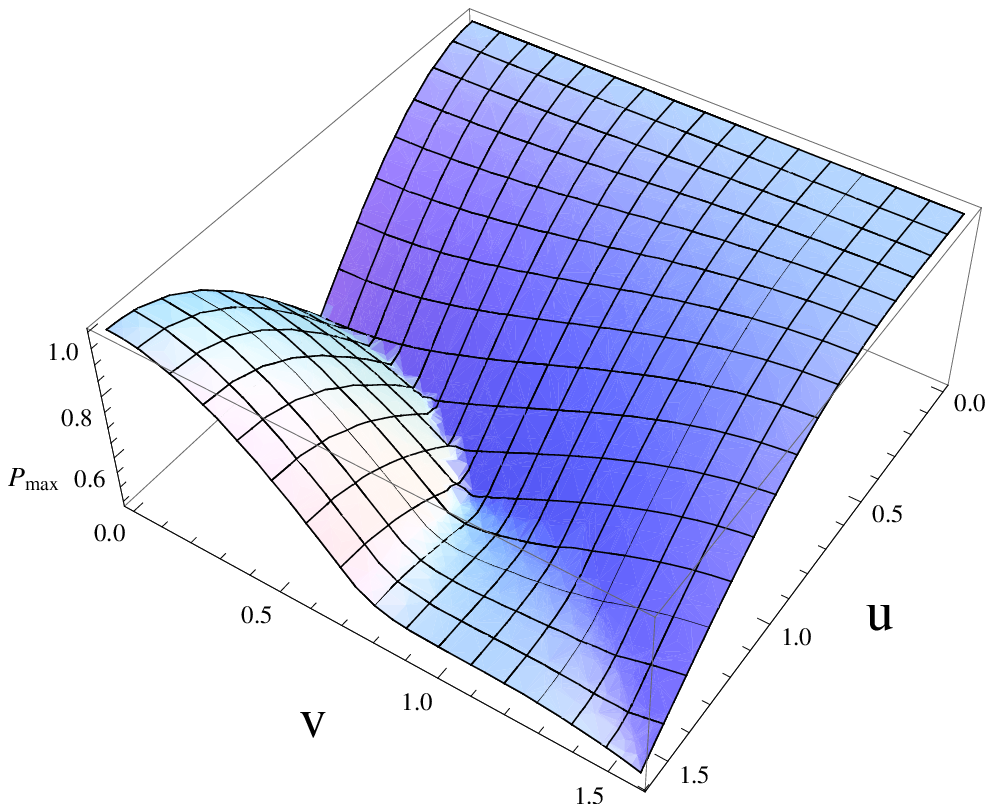}
\includegraphics[height=6cm]{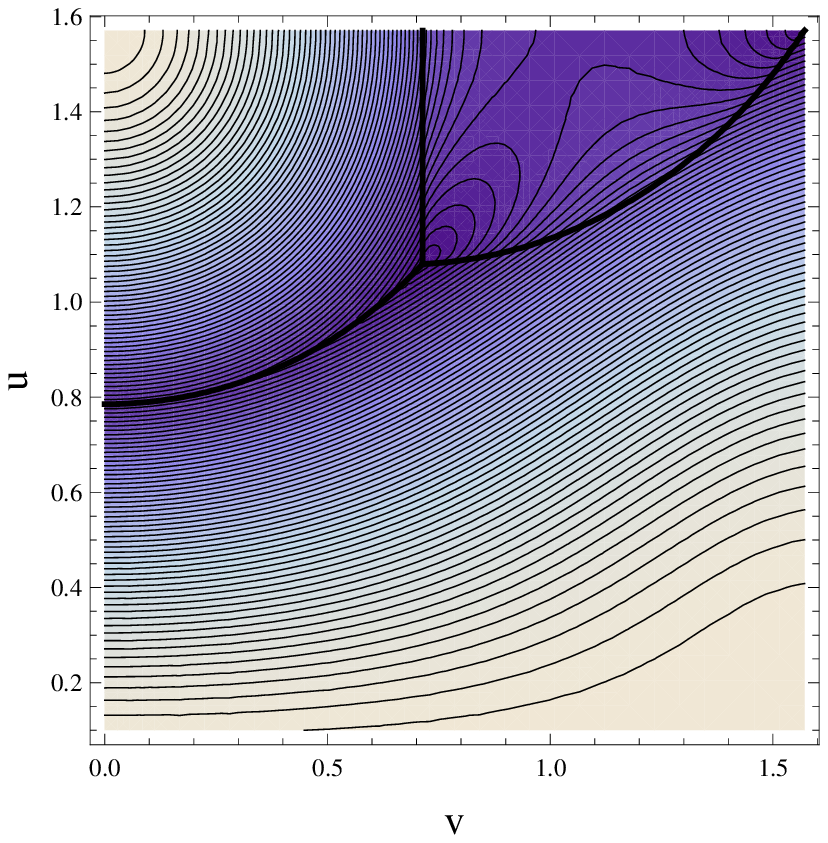}
\caption[fig4]{(Color online) Fig. 4(a) is a plot of the applicable domains for $\gamma=\pi/2$
case in $(u,v)$-plane. Unlike $\gamma=0$ case there are three applicable domains in this 
case. The principal domain $P_{max} = \nu_P^2$ is larger than $P_{max} = \mu_P^2$ in 
$\gamma=0$ case. This fact seems to indicate that the principal domain increases its 
territory with increasing $\gamma$. It is important to note that the domain $P_{max} = \nu_+^2$
is not reached to $h=0$ axis. This implies the consistency of the $h\rightarrow 0$ limit. 
Fig. 4(b) is $(u,v)$-dependence of $P_{max}$. The highly entangled states forms a valley
between two mountains. Fig. 4(c) is a plot of $P_{max}$ and the applicable domains in the 
$(u,v)$-plane. The boundaries of the domains are represented by black think line. Many 
highly-entangles states reside around the boundaries and in the domain $P_{max} = \nu_2^2$. It 
is mainly due to the fact that there are two LU-equivalent W-states for $\gamma=\pi/2$ case.}
\end{center}
\end{figure}

Unlike $\gamma = 0$ case the whole parameter space is divided into the three applicable
domains. Introducing the parameters $u$ and $v$ as Eq.(\ref{4-para-1}) we plot the three
applicable domains in the $u$-$v$ plane in Fig. 4a. Around $h=0$ axis there are two
domains, {\it i.e.} $\nu_P^2$ and $\nu_2^2$. Since $\nu_P^2$ and $\nu_2^2$ go to
$\mu_P^2$ and $\mu_+^2$ in the $h \rightarrow 0$ limit, this guarantees that the
$h \rightarrow 0$ limit is consistent with same limit of $\gamma = 0$ case. The applicable
domain for $\nu_P^2$ is little bit larger than the domain $\mu_P^2$ for $\gamma=0$ case.
The point ($u=\cos^{-1}(\sqrt{2}/3), v=\tan^{-1}(\sqrt{3}/2)$) is shared by three domains.
This point corresponds to
\begin{equation}
\label{5-w-1}
|\psi_W \ra = \frac{2}{3} |000\ra + \frac{1}{3} \left(|011\ra + |101\ra + |110\ra \right)
+ i \frac{\sqrt{2}}{3} |111\ra.
\end{equation}
This is LU-equivalent with $|W\ra = (1/\sqrt{3})(|100\ra + |010\ra + |001\ra)$
as shown in Ref.\cite{maximal}.

In Fig. 4b we plot the ($u,v$)-dependence of $P_{max}$ given in Eq.(\ref{5-final}). Like
Fig. 1b the highly entangled states are represented as a valley in this figure. Fig. 4b
seems to show that there exists an alley in the valley, which ends at $u=v=\pi/2$. Along this
alley so many highly entangled states are located. Comparing Fig. 4b with Fig. 1b, one
can realize that there are many more highly-entangles states for $\gamma=\pi/2$ case than
$\gamma=0$ case. This is mainly due to the fact that there are two LU-equivalent
W-states when $\gamma = \pi/2$.

Fig. 4c shows the geometric entanglement measure and the applicable domains simultaneously
in the $u$-$v$ plane. Fig. 4c shows that around two W-states there are so many highly
entangled states, which we would like to call W-neighbors. Especially, the neighbors of
$|\psi_W\ra$ in Eq.(\ref{5-w-1}) gather along ${\cal C}_3=0$ line. Besides the W-neighbors
there are many highly entangled states around boundary of the applicable domains. These are
the neighbors of the shared states\cite{shared}, and we would like to call them
the GHZ-neighbors. The GHZ-neighbors are slightly less-entangled compared to the W-neighbors.
However, the number of the GHZ-neighbors are many more than that of the W-neighbors.

Finally, we consider the several special cases. First example is $h=0$ case. In this case
${\cal C}_2 = -4g^2t \leq 0$ and ${\cal C}_3 = -(g-t)^2 (g+2t) \leq 0$, which results in
identical expression with Eq.(\ref{4-ex-2}). Therefore, both results for $\gamma=0$ and
$\gamma=\pi/2$ cases coincide with each other in the $h \rightarrow 0$ limit. Second example
is $t=0$ case. It is easy to show that in this case $P_{max}=g^2$ when $g \geq h$ and
$P_{max}=h^2$ when $g\leq h$. This is consistent with
$P_{nax}(GHZ) = \max (|\alpha|^2, |\beta|^2)$ when $|GHZ \ra = \alpha |000\ra + |111\ra$.

\section{Eigenvalues and Geometric measure for $\gamma = \pi/4$ : Numerical Approach}

In this section we will compute the eigenvalues and the geometric measure for
$\gamma = \pi/4$ case.

\subsection{Eigenvalues}

For $\gamma = \pi/4$ Eq.(\ref{main-2}) reduces to
\begin{subequations}
\label{ga-pi-4}
\begin{equation}
\label{ga-pi-4x}
2t(g+t)\sin\theta\cos\varphi+\sqrt{2}ht(1-\cos\theta)=\lambda\sin\theta\cos\varphi
\end{equation}
\begin{equation}
\label{ga-pi-4y}
- 2t(g-t)\sin\theta\sin\varphi+\sqrt{2}ht(1-\cos\theta)=\lambda\sin\theta\sin\varphi
\end{equation}
\begin{equation}
\label{ga-pi-4z}
(g^2-t^2)(1+\cos\theta)-h^2(1-\cos\theta)-\sqrt{2}ht\sin\theta(\sin\varphi+\cos\varphi)=\lambda
\cos\theta.
\end{equation}
\end{subequations}

When $\theta=0$, Eq.(\ref{ga-pi-4x}) and Eq.(\ref{ga-pi-4y}) are automatically solved and
Eq.(\ref{ga-pi-4z}) gives
\begin{equation}
\label{6-eigen-1}
\lambda=2(g^2-t^2).
\end{equation}
Since ${\bm s} = (0,0,1)$ for this case, from Eq.(\ref{eigenvalue-1}) the corresponding
eigenvalue is
\begin{equation}
\label{6-eigen-2}
\rho_P^2 = g^2.
\end{equation}

When $\sin\theta\neq0$, Eq.(\ref{ga-pi-4x}) and Eq.(\ref{ga-pi-4y}) reduce to
\begin{equation}
\label{6-z-1}
z=\frac{\lambda-2gt-2t^2}{\sqrt{2}ht}\cos\varphi=\frac{\lambda+2gt-2t^2}{\sqrt{2}ht}\sin\varphi
\end{equation}
where $z=\tan(\theta/2)$. From Eq.(\ref{6-z-1}) one can compute $\varphi$ if $\lambda$ is
known by using
\begin{equation}
\label{6-phi-1}
\tan\varphi=\frac{(\lambda-2t^2)-2gt}{(\lambda-2t^2)+2gt}.
\end{equation}
Deriving $\sin\varphi+\cos\varphi$ from Eq.(\ref{6-z-1}) and inserting it into
Eq.(\ref{ga-pi-4z}), one can derive the expression of $z^2$ in a form
\begin{equation}
\label{6-z-2}
z^2=\frac{\left[(\lambda-2t^2)^2-4g^2t^2\right]
                (\lambda-2g^2+2t^2)}{(\lambda-2h^2)(\lambda-2t^2)^2-
                          8h^2t^2(\lambda-2t^2)-4g^2t^2(\lambda-2h^2)}.
\end{equation}
On the other hand, one can derive a different expression of $z^2$ directly from
Eq.(\ref{6-z-1})
\begin{equation}
\label{6-z-3}
z^2 = \frac{(\lambda-2gt-2t^2)^2}{2h^2t^2} (1+\tan^2\varphi)^{-1} =
\frac{\left[(\lambda-2t^2)^2-4g^2t^2\right]^2}{4h^2t^2\left[(\lambda-2t^2)^2+4g^2t^2\right]}.
\end{equation}
Equating Eq.(\ref{6-z-2}) with Eq.(\ref{6-z-3}) yields an equation for solely $\lambda$:
\begin{equation}
\label{6-lambda-1}
\lambda f(\lambda) = 0
\end{equation}
where
\begin{eqnarray}
\label{6-lambda-2}
& & f(\lambda) = \lambda^4-2(h^2+4t^2)\lambda^3 - 4t^2(2g^2-h^2-6t^2)\lambda^2  \\  \nonumber
& & \hspace{1.5cm}  +8\left[t^4(h^2-4t^2)+g^2(3h^2t^2+4t^4)\right]\lambda    \\   \nonumber
& & \hspace{1.5cm}  +   16t^4\left(g^4-5g^2h^2-2g^2t^2-h^2t^2+t^4\right).
\end{eqnarray}
Eq.(\ref{6-lambda-1}) guarantees the existence of the eigenvalue for $\lambda=0$ as
$\gamma=0$ and $\gamma=\pi/2$ cases. In fact, one can show that there exists an eigenvalue
corresponding to $\lambda=0$ for arbitrary $\gamma$. We have shown this fact in
appendix A.

When $\lambda=0$, Eq.(\ref{6-phi-1}) and Eq.(\ref{6-z-3}) reduce to
\begin{equation}
\label{6-z-4}
z^2 = \frac{(g^2-t^2)^2}{h^2(g^2+t^2)} \hspace{1.0cm} \tan \varphi=-\frac{g+t}{g-t}.
\end{equation}
Combining Eq.(\ref{6-z-1}) and Eq.(\ref{6-z-4}), the possible solutions for $\theta$ and
$\varphi$ are
\begin{equation}
\label{6-z-5}
z=\pm\frac{g^2-t^2}{h\sqrt{g^2+t^2}} \hspace{1.0cm}
\cos\varphi=\mp\frac{g-t}{\sqrt{2(g^2+t^2)}} \hspace{1.0cm}
\sin\varphi=\pm\frac{g+t}{\sqrt{2(g^2+t^2)}}.
\end{equation}
It is easy to show that both solutions in Eq.(\ref{6-z-5}) gives a same eigenvalue, which
is
\begin{equation}
\label{6-eigen-3}
\rho_0^2 = \frac{g^2(g^2+t^2)h^2 + t^2(g^2-t^2)^2}{h^2(g^2+t^2)+(g^2-t^2)^2}.
\end{equation}

Finally, let us consider $f(\lambda)=0$. It is worthwhile noting that at $h\rightarrow 0$ limit
$f(\lambda)=0$ reduces to $(\lambda-2gt-2t^2)^2(\lambda+2gt-2t^2)^2=0$. Therefore, the
eigenvalues corresponding to $f(\lambda)=0$ should coincide with $\mu_{\pm}^2$ and $\mu_2^2$
for $\gamma=0$ case, and with $\nu_{\pm}^2$ and $\nu_2^2$ for $\gamma=\pi/2$ case at the
$h\rightarrow 0$ limit. Equation  $f(\lambda)=0$ gives four solutions of $\lambda$, {\it say}
$\lambda_1$, $\lambda_2$, $\lambda_3$ and $\lambda_4$. We ordered the solutions by a fact that
the $h\rightarrow 0$ limit
of $\lambda_1$ and $\lambda_2$ is $-2t(g-t)$ and same limit of $\lambda_3$ and $\lambda_4$ is
$2t(g+t)$. Then, the corresponding eigenvalues, {\it say} $\rho_1^2$, $\rho_2^2$, $\rho_3^2$,
and $\rho_4^2$, can be computed numerically.

\subsection{geometric measure}

\begin{figure}[ht!]
\begin{center}
\caption[fig4]{(Color online) Fig. 5(a) is a plot of the applicable domains for $\gamma=\pi/4$
case. In this case there are two applicable domains. The principal domain $P_{max} = \rho_P^2$
is little bit larger than $P_{max}=\mu_P^2$ for $\gamma=0$ and little bit smaller than
$P_{max}=\nu_P^2$ for $\gamma=\pi/2$. This fact indicates that the principal domain increases
its territory with increasing $\gamma$. Fig. 5(b) is $(u,v)$-dependence of $P_{max}$. As 
$\gamma=0$ case the highly-entangled states form a valley between two mountains. Fig. 5(c) is
a plot of $P_{max}$ and the applicable domains in the $(u,v)$-plane. Many highly-entangled 
states reside around boundary of the domains and near W-state.}
\end{center}
\end{figure}

Using eigenvalues $\rho_P^2$, $\rho_0^2$ derived analytically and
$\rho_i^2 \hspace{.2cm} (i=1,2,3,4)$
computed numerically, one can compute $P_{max}$ for the $\gamma=\pi/4$ case. Since each
eigenvalue has its own available region, we checked this region by imposing
$\mbox{Re}[\lambda] = 0$, $-1 \leq \sin\theta \leq 1$, $-1 \leq \cos\theta \leq 1$,
$-1 \leq \sin\varphi \leq 1$, and $-1 \leq \cos\varphi \leq 1$. Although there are six
different eigenvalues, the numerical calculation shows that only $\rho_P^2$ and $\rho_4^2$
contribute to the geometric measure. This indicates that the whole parameter space is divided
into two applicable domains. These two domains are represented in $u-v$ plane in Fig. 5a. The
domains $\rho_P^2$ is slightly larger than domain $\mu_P^2$ and slightly smaller than
domain $\nu_P^2$. This fact seems to indicate that the domain containing $g=1$ extends its
territory with increasing $\gamma$.

Fig. 5b is a $(u,v)$-dependence of $P_{max}$ for $\gamma = \pi/4$. Similarly with
$\gamma = 0$ and $\pi/2$ cases, many highly entangled states reside at the valley between two
mountains. Another highly entangled states reside around $u=v=\pi/2$, which corresponds to
W-state. The alley appeared in Fig. 4b does not appear in this case. This seems to be due to
the fact that there is only one W-state in $\gamma = \pi/4$ case.

Fig. 5c is a $(u,v)$-dependence of $P_{max}$ and domains. As expected the highly entangled
states are located around boundary and W-state.

\section{Conclusion}
In this paper we have explored the effect of the phase factor in the geometric entanglement
measure. We have chosen the most general three-qubit states which have symmetry under the
qubit-exchange. Our choice of the quantum states enables us to derive all eigenvalues and
geometric measure analytically when the phase factor $\gamma$ is $0$ or $\pi/2$. It turns out
that the $\gamma=\pi/2$ case has three applicable domains while the $\gamma=0$ case has two
domains. Most highly entangled states reside around the boundaries of the domains and near
W-state. Apart from the boundaries more and more the quantum states lose their entanglement
and eventually, become the product states.

Our result naturally gives rise to a question:
what is a critical $\gamma$, {\it say} $\gamma_c$,
which distinguish the two and three domains? In order to explore this question we have
analyzed the $\gamma = \pi/4$ case numerically. Our numerical calculation shows that there
are six different eigenvalues for $\gamma = \pi/4$ case, but only two of them contribute to
the geometric entanglement measure. Thus, there are two domains for $\gamma = \pi/4$.

We conjecture that emergence of the three applicable domains at $\gamma=\pi/2$ is due to
the two LU-equivalent W-states. In order to confirm our conjecture we checked numerically
$\gamma=\pi/3$ and $\gamma=11 \pi/24$ cases, which also give two applicable domains. We also
checked the applicable domains for the partially symmetric quantum state
\begin{equation}
\label{partial}
|\psi\rangle=g|000\rangle + t|011\rangle + t|101\rangle + t_3 |110\rangle + e^{i\gamma}
h|111\rangle
\end{equation}
numerically when $\gamma=0$. This case also gives two applicable domains. Therefore, we 
conclude that the emergence of the three applicable domains is due to the appearance of 
additional W-state.

In appendix we have shown that there exist eigenvalues for all $\gamma$, whose Lagrangian
multiplier constant is zero. Although we conjecture that this is due to some symmetry of
the quantum state $|\psi\rangle$, we do not know the exact physical reason for the
emergence of these solutions. It seems to be of interest to reveal the physical meaning
of these solutions clearly.

{\bf Acknowledgement}:
This work was supported by National Research Foundation of Korea Grant funded by the
Korean Government (2009-0073997).

\newpage

\begin{appendix}
{\centerline{\bf Appendix A}}
\setcounter{equation}{0}
\renewcommand{\theequation}{A.\arabic{equation}}

In this appendix we would like to show the existence of the eigenvalue $\mu_0^2$, which
corresponds to $\lambda = 0$, at arbitrary $\gamma$. When $\lambda=0$, Eq.(\ref{main-2})
reduces to
\begin{subequations}
\label{ga-appen}
\begin{equation}
\label{ga-appen-x}
2ht \cos\gamma(1-\cos\theta)+2t(g+t)\sin\theta\cos\varphi=0
\end{equation}
\begin{equation}
\label{ga-appen-y}
2ht\sin\gamma(1-\cos\theta)-2t(g-t)\sin\theta\sin\varphi=0
\end{equation}
\begin{equation}
\label{ga-appen-z}
(g^2-t^2)(1+\cos\theta)-h^2(1-\cos\theta)-2ht\sin\theta\cos(\varphi-\gamma)=0.
\end{equation}
\end{subequations}
The existence of $\mu_0^2$ can be shown as following. First we derive $\theta$ and $\varphi$ by
making use of Eq.(\ref{ga-appen-x}) and Eq.(\ref{ga-appen-y}). Then we show that the
solutions $\theta$ and $\phi$ also solve Eq.(\ref{ga-appen-z}).

Now, we consider only $\sin\theta\neq 0$ case. Then from Eq.(\ref{ga-appen-x}) and
Eq.(\ref{ga-appen-y}) it is easy to derive
\begin{equation}
\label{appen-2}
(g+t)\sin\gamma\cos\varphi + (g-t)\cos\gamma\sin\varphi=0,
\end{equation}
which gives
\begin{equation}
\label{appen-3}
\tan\varphi=-\frac{g+t}{g-t}\tan\gamma.
\end{equation}
Combining Eq.(\ref{appen-2}) and Eq.(\ref{appen-3}), one can derive the solution for $\varphi$,
which is
\begin{equation}
\label{appen-4}
\cos\varphi=\pm\frac{g-t}{\sqrt{(g-t)^2+(g+t)^2\tan^2\gamma}}  \hspace{1.0cm}
\sin\varphi=\mp\frac{(g+t)\tan\gamma}{\sqrt{(g-t)^2+(g+t)^2\tan^2\gamma}}.
\end{equation}
Inserting Eq.(\ref{appen-4}) into Eq.(\ref{ga-appen-y}), one can derive $\sin\theta$ in
a form
\begin{equation}
\label{appen-5}
\sin\theta=\mp\frac{2h(g^2-t^2)\sqrt{g^2+t^2-2gt\cos2\gamma}}{h^2(g^2+t^2-2gt\cos2\gamma) +
                                                              (g^2-t^2)^2}.
\end{equation}
Inserting Eq.(\ref{appen-4}) and Eq.(\ref{appen-5}) into the lhs of Eq.(\ref{ga-appen-z}), one
can show straightforwardly that Eq.(\ref{ga-appen-z}) is solved already by Eq.(\ref{appen-4})
and Eq.(\ref{appen-5}). This guarantees the existence of $\mu_0^2$.

In order to derive $\mu_0^2$ explicitly we choose the upper sign in Eq.(\ref{appen-4}) and
Eq.(\ref{appen-5}). Then the components of the vector ${\bm s}$ becomes
\begin{subequations}
\label{appen-6}
\begin{equation}
\label{appen-6-x}
s_x = \sin\theta\cos\varphi=\frac{-2h(g-t)^2(g+t)\cos\gamma}
                                  {h^2 [(g^2+t^2)-2gt\cos2\gamma] + (g^2-t^2)^2}
\end{equation}
\begin{equation}
\label{appen-6-y}
s_y=\sin\theta\sin\varphi=\frac{2h(g-t)(g+t)^2\sin\gamma}
                               {h^2 [(g^2+t^2)-2gt\cos2\gamma] + (g^2-t^2)^2}
\end{equation}
\begin{equation}
\label{appen-6-z}
s_z = \cos\theta=\frac{h^2 [(g^2+t^2)-2gt\cos2\gamma] - (g^2-t^2)^2}
                      {h^2 [(g^2+t^2)-2gt\cos2\gamma] + (g^2-t^2)^2}.
\end{equation}
\end{subequations}
Inserting Eq.(\ref{appen-6}) into Eq.(\ref{eigenvalue-1}) and performing tedious calculation,
one can show that $\mu_0^2$, eigenvalue corresponding to $\lambda=0$, becomes
\begin{equation}
\label{appen-7}
\mu_0^2=\frac{g^2h^2(g^2+t^2-2gt\cos2\gamma)+t^2(g^2-t^2)^2}
             {h^2(g^2+t^2-2gt\cos2\gamma)+(g^2-t^2)^2}.
\end{equation}
It is straightforward to show that the choice of lower sign in Eq.(\ref{appen-4}) and
Eq.(\ref{appen-5}) leads us to same expression of $\mu_0^2$.
One can show easily that $\mu_0^2$ exactly coincides with $\mu_1^2$ in Eq.(\ref{ev-0-2}),
$\nu_1^2$ in Eq.(\ref{ev-1-2}) and $\rho_0^2$ in Eq.(\ref{6-eigen-3}) when $\gamma=0$,
$\gamma=\pi/2$ and $\gamma=\pi/4$ respectively.

Finally, making use of explicit expression of $\mu_0^2$, oen can derive the nearest product
state $|q\rangle|q\rangle|q'\rangle$ for $\mu_0^2$, i.e.
\begin{equation}
\label{appen-8}
_{AB}\langle q|\langle q|\psi\rangle = \mu_0 |q'\rangle    \hspace{1.0cm}
_{AC}\langle q|\langle q'|\psi\rangle = \mu_0 |q\rangle    \hspace{1.0cm}
_{BC}\langle q|\langle q'|\psi\rangle = \mu_0 |q\rangle    \hspace{1.0cm}
\end{equation}
where $|\psi\rangle$ is given in Eq.(\ref{symmetric}). Since ${\bm s}$ is a Bloch vector
of $|q\rangle \langle q|$, one can show directly
\begin{equation}
\label{appen-9}
|q\rangle = \frac{1}{\sqrt{h^2\ell^2 + (g^2-t^2)^2}}
            \left[h\ell |0\rangle - (g^2-t^2) e^{-i\eta} |1\rangle \right]
\end{equation}
where
\begin{equation}
\label{appen-10}
\ell^2\equiv g^2+t^2-2gt\cos2\gamma  \hspace{1.0cm}
\cos\eta=\frac{g-t}{\ell}\cos\gamma  \hspace{1.0cm}
\sin\eta=\frac{g+t}{\ell}\sin\gamma.
\end{equation}
Inserting Eq.(\ref{appen-9}) into Eq.(\ref{appen-8}) it is straightforward to show
that $|q'\rangle$
becomes
\begin{equation}
\label{appen-11}
|q'\rangle = \frac{1}{{\cal N}} \left[\left\{gh^2\ell^2+t(g^2-t^2)^2e^{2i\eta}\right\}|0\rangle
+ e^{i\eta}h(g^2-t^2)\left\{(g^2-t^2)e^{i(\gamma+\eta)}-2\ell t\right\}|1\rangle\right]
\end{equation}
where ${\cal N}$ is a normalization constant, which makes $|q'\rangle$ unit vector.

For $\gamma=0$ case the nearest product state becomes
\begin{eqnarray}
\label{appen-12}
& &|q\rangle = \frac{1}{\sqrt{h^2+(g+t)^2}} \left(h|0\rangle-(g+t)|1\rangle \right)
                                                                     \\   \nonumber
& &|q'\rangle=\frac{1}{\sqrt{\left\{gh^2+t(g+t)^2\right\}^2+h^2(g^2-t^2)^2}}
   \left[\left\{gh^2+t(g+t)^2\right\}|0\rangle+h(g^2-t^2)|1\rangle\right].
\end{eqnarray}
It is interesting to note that $\langle q|q'\rangle=0$ when ${\cal D}_1 = 0$, where
${\cal D}_1$ is given in Eq.(\ref{4-cri-2}).

For $\gamma=\pi/2$ case $|q\rangle$ abd $|q'\rangle$ becomes
\begin{eqnarray}
\label{appen-13}
& &|q\rangle = \frac{1}{\sqrt{h^2+(g-t)^2}} \left(h|0\rangle+ i(g-t)|1\rangle \right)
                                                                     \\   \nonumber
& &|q'\rangle=\frac{1}{\sqrt{\left\{gh^2-t(g-t)^2\right\}^2+h^2(g^2-t^2)^2}}
   \left[\left\{gh^2-t(g-t)^2\right\}|0\rangle- i h(g^2-t^2)|1\rangle\right].
\end{eqnarray}
It is interesting to note that $\langle q|q'\rangle=0$ when ${\cal C}_3 = 0$, where
${\cal C}_3$ is given in Eq.(\ref{5-cri-3}).

\end{appendix}

\end{document}